\documentclass[pra,aps,amsfonts]{revtex4} 
\usepackage{amsthm}
\usepackage{epsfig}
\usepackage{graphicx}

\newcommand{\beq}{\begin{equation}}
\newcommand{\eeq}{\end{equation}}
\newcommand{\bei}{\begin{itemize}}
\newcommand{\eei}{\end{itemize}}
\newcommand{\bee}{\begin{enumerate}}
\newcommand{\eee}{\end{enumerate}}

\newcommand{\beqa}{\begin{eqnarray}}
\newcommand{\eeqa}{\end{eqnarray}}
\newcommand{\beqar}{\begin{eqnarray*}}
\newcommand{\eeqar}{\end{eqnarray*}}

\def\be{\begin{equation}}
\def\ee{\end{equation}}
\def\ben{\begin{eqnarray}}
\def\een{\end{eqnarray}}

\def\eea{\end{array}}
\def\bea{\begin{array}}

\newcommand{\tr}{{\rm tr}}

\newtheorem{theorem}{Theorem}

\newtheorem{definition}{Definition}
\newtheorem{lemma}{Lemma}
\newtheorem{proposition}{Proposition}

\def\ot{\otimes}

\def \D {{\Delta }}
\def \Dhv {{C_{HV}}}
\def \Dcl {{\Delta_{cl}}}
\def \Dc {{\Delta_{cl}}}
\def \hcal {{\cal H}}
\def \pcal {{\cal P}}
\def \ccal {{\cal C}}

\def \s {\,\,\,\,}

\def \both {{\leftrightarrow}}

\def \o{\emptyset}

\def \r {\varrho}
\def \rab {\varrho_{AB}}
\def \ILO {I_{LO}} 

\def \Il {I_l}
\def\I{I}

\def \s {\,\,\,\,}
\def\rab{\rho_{AB}}
\def\vr{\varrho}
\def\D{\Delta}
\def\Dc{\Delta_c}
\def\r{\rho}
\def\N{N}
\def\rn{\vr}

\def\SA{S(\rho_A)}
\def\SB{S(\rho_B)}
\def\SAB{S(\rho_{AB})}
\def\Im{I_M}
\def\Er{E_r}

\def\textbf#1{{\bf #1}}

\def\>{\rangle}
\def\<{\langle}
\def\blacksquare{\vrule height 4pt width 3pt depth2pt}
\def\rab{\varrho_{AB}}

\def\clcor{classically correlated}
\def\pclcor{pseudo-classically correlated}
\def\pc{\pcal\ccal}
\def\reldist{E^{\pc}_r}

\def\duzomniejsze{<\kern-.7mm<}
\def\duzowieksze{>\kern-.7mm>}

\def\textbf#1{{\bf #1}}

\def\blacksquare{\vrule height 4pt width 3pt depth2pt}

\def\pcal{{\cal P}}
\def\hcal{{\cal H}}

\def\bcal{{\cal B}}
\def\scal{{\cal S}}
\def\rab{\varrho_{AB}}
\def\r{\varrho}

\def\dt#1{{{\kern -.0mm\rm d}}#1\,}
\def\ypodpis{\raise4mm\hbox{$\omega$}}

\def \s {\,\,\,\,}
\def \singlet {\psi^-}
\newcommand{\braket}[1]{\ket{#1}\bra{#1}}
\newcommand{\bra}[1]{\langle #1 |}
\newcommand{\ket}[1]{| #1 \rangle}
\def\ers{E_r^S}

\def\textbf#1{{\bf #1}}

\def\>{\rangle}
\def\<{\langle}
\def\blacksquare{\vrule height 4pt width 3pt depth2pt}

\begin{document}

\title{Local versus non-local information in 
quantum information theory: formalism and phenomena }

\author{Micha\l{} Horodecki$^{(1)}$,  Pawe\l{}
Horodecki$^{(2)}$, Ryszard Horodecki$^{(1)}$,
Jonathan Oppenheim$^{(1)(3)(4)}$, Aditi Sen(De)$^{(1),(5)}$, 
Ujjwal Sen$^{(1),(5)}$ and Barbara Synak\(^{(1)}\)}

\affiliation{$^{(1)}$Institute of Theoretical Physics and Astrophysics,
University of Gda\'nsk, 80--952 Gda\'nsk, Poland}
\affiliation{$^{(2)}$Faculty of Applied Physics and Mathematics,
Technical University of Gda\'nsk, 80--952 Gda\'nsk, Poland}
\affiliation{$^{(3)}$ Dept. of Applied Mathematics and Theoretical Physics, University of
Cambridge, Cambridge, U.K.}
\affiliation{$^{(4)}$Racah Institute of Theoretical Physics, 
Hebrew University of Jerusalem, Givat Ram, Jerusalem 91904, Israel}
\affiliation{\(^{(5)}\)Institut f\"ur Theoretische Physik, Universit\"at Hannover, D-30167 Hannover,
Germany}
\date{October 4th, 2004}

\begin{abstract}

In spite of many results in quantum information theory,
the complex nature of compound systems is far from being clear. In general 
the information is a mixture of local, and non-local ("quantum")
information. It is important from both pragmatic and theoretical points of view 
to know relationships between the two components. To make this point more clear,
we develop and investigate the quantum information processing paradigm 
in which parties sharing a multipartite state distill {\it local}
information. The amount of information which is lost because the parties must use a
classical communication channel is the {\it deficit}. 
 This scheme can be viewed as {\it complementary} to the notion of distilling
entanglement.
 After reviewing the paradigm in detail, we show that
the upper bound for the deficit is given by the relative entropy distance 
to so-called \pclcor\ states; the lower bound is 
the relative entropy of entanglement. 
This implies, in particular, that 
{\it any entangled state is informationally nonlocal} i.e. 
has nonzero deficit.  We also apply the paradigm 
to defining the thermodynamical cost of erasing entanglement. 
We show the cost is bounded from below by relative 
entropy of entanglement.
We demonstrate the existence of several other non-local phenomena 
which can be found using the paradigm of local information. 
For example,
we prove the existence of a form of non-locality without entanglement and with
distinguishability. 
We analyze the deficit for several classes of multipartite pure states 
and obtain that in contrast to the GHZ state, the Aharonov state 
is extremely nonlocal (and in fact can be thought of as quasi-nonlocalisable). 
We also show that there 
do not exist states, for which the deficit is strictly equal to the whole informational 
content ({\it bound local information}). We discuss the relation of the paradigm with measures 
of classical correlations introduced earlier. It is also proven
that in the one-way scenario, the deficit is additive for 
Bell diagonal states.  We then discuss complementary features 
of information in distributed quantum systems. 
Finally we discuss the physical and theoretical 
meaning of the results and pose many open questions.

\end{abstract}
\maketitle
\tableofcontents

\section{Introduction}
\label{intro}

"Quantum information" is emerging as a primitive notion in physics following
an essential extension of classical 
Shannon information theory \cite{Ingarden-qit} into the quantum domain. 
Quantum information can not be defined precisely, but it is necessary to understand the role
of this mysterious
and "unspeakable" information \cite{Peres-IBM} in newly discovered quantum phenomena 
such as teleportation  \cite{teleportation}
or cryptography \cite{BB84}\cite{ekert-qkd}.
These phenomena strongly suggest that quantum states 
represent quantum information - reality we process in the laboratory, but which can not 
be described as a sequence  of classical symbols on 
a Turing tape \cite{HHH-IBM,Jozsa-IBM}. Recently the 
no-deleting and no-cloning theorems  have been connected with  the principle 
of conservation of quantum information \cite{HHSS-clo-del}. Like physical quantities such as 
energy, quantum information has  different forms 
and one of them is entanglement - an exotic resource extraordinarily sensitive to the environment.
One finds a loss of entanglement in the transition from pure entangled state to noisy entangled state, yet
remarkably this process can be partially reversed within the distance labs paradigm. 
Namely from a large number of noisy bipartite states shared between two distant parties one can distill a number of
e-bits at the optimal conversion rate using local operations and classical communications (LOCC) \cite{BBPSSW1996}.

Despite a plethora of measures which can be used to quantify
entanglement, we are still far from properly understanding it.
Part of the difficulty is that measures of a quantity are
not enough to understand the quantity - one needs to understand
entanglement in relation to something else.  You cannot understand
entanglement in relation to entanglement.     
In the above context, basic questions arise:
{\it i) Does entanglement exhaust all aspects of quantum information?
ii) Are there resources other than entanglement in the distant labs 
paradigm? iii) Does quantum information involve a nonlocality which goes beyond Bell theorem?}

The above questions have been recently 
considered 
\cite{Bennett-nlwe,Zurek-discord,Zurek01-discord,
HendersonVedral,OHHH2001,nlocc,compl,phase,Vedral03-thermo-ent,SynakHH04,igor-deficit}. 
In particular,a new quantum information processing 
paradigm has been introduced,  where we proposed the idea of attributing cost to  local
resources such as pure local qubits \cite{OHHH2001,nlocc}.  
Instead of asking how  much entanglement can be distilled from a state 
shared between two parties,  one can ask how many local pure qubits 
$\Il$ can be drawn from it.
This gives a quantity (called {\it localisable information})
which can then be used to get insight into the double nature of quantum information.
Namely, it was shown that local information can be thought of as 
being complementary to entanglement  \cite{compl}, thereby allowing one, 
in particular, to understand entanglement in relation to
$\Il$.

At first glance, the idea of considering local pure states
to be a resource may seem curious.  In traditional entanglement
theory, one thinks of local pure states as being a free resource.
Each party can use as many pure state ancillas as desired.  
Furthermore, one can obtain pure states from a mixed state, simply
by performing a measurement on the state.  
Note however, that the second law of thermodynamics 
tells us that purity is indeed a resource.
One can never decrease the entropy of a closed system; entropy only
increases. The reason a measurement appears to produce pure states is that
we ignore the fact that the measuring apparatus must have initially
been set in some pure state, and after the measurement, the 
apparatus will be in a mixture of all the different measurement outcomes.
In other words, in a closed system which includes the state, the 
measurement apparatus and the observer, the total number of pure qubits
can never increase.  We must therefore be careful how
we define the allowable class of operations in order to account for
all pure states which might be introduced by various parties from
the outside.  We will discuss such a useful class, called
{\it Closed Operations}  which can 
properly be used to account for pure states.

By considering pure states as a resource, one is immediately 
connecting quantum information theory with thermodynamics.
In fact, it was the early foundational
work on reversible computation \cite{Bennett82} 
where the entropic cost of computation was considered \cite{Landauer}.  This became
one of the cornerstones which led to the possibility of quantum computation.  
The relationship between information and physical tasks such as performing work, also has
a long history beginning with Szilard \cite{szilard}.
In fact, as shown in \cite{nlocc,uniqueinfo} the information function 
is exactly equal to the number of pure qubits one
can extract from a state while having many copies of the state. We will therefore talk of extracting
{\it information} $\I$ from a state. One can think of this as
extracting pure states from more mixed states.   From the work of Szilard,
we also know that the information $\I$ is closely related to
the amount of work one can extract from a single heat bath (see \cite{silnik} for a rigorous 
derivation). These connections will be discussed
in Section \ref{sec:information} were we review the basic concepts.

The rough essence of the approach is that if separated individuals
extract local pure states (i.e. information) from a shared state, 
using only local operations and classical communication,
then they will in general be able to extract less information than
if they were together. If the amount of information they can
extract when they are together from a state $\varrho$ 
is $\I(\varrho)$, and the optimal \cite{twist} 
amount they
can extract when separated is $\Il(\varrho)$, then the difference
(called the {\it deficit}) $\D(\varrho)\equiv
I(\varrho)-\Il(\varrho)$
feels some non-classical correlations in the state $\varrho$.
or

Note that the quantity $\D$ is not an entanglement measure, at least
in the regime of finite copies of a state $\r$. It feels not only entanglement,
but also, so called {\it non-locality without entanglement}
\cite{Bennett-nlwe}.  We say that it quantifies {\it the quantumness of
correlations} rather than entanglement (first attempts to formally quantify such features for quantum states are due to 
\cite{Zurek-discord}, and for ensembles in \cite{Bennett-nlwe}). The state which has nonzero deficit 
we will call "informationally nonlocal". The term {\it nonlocality} means here 
that distant parties can do worse than parties that are together, despite the fact that they
can communicate classically \cite{Bennett-nlwe}. Thus it is a different notion than 
the nonlocality understood as a violation of local realism (we have discussed the relations 
in \cite{epr-nlwe}).

In this work we review some of the results of \cite{OHHH2001,compl, nlocc, phase, uniqueinfo}
and provide more detail.  We then give
a number of new, essential results within
the paradigm of distillation of local information. 
In particular we provide a lower bound for the deficit:
it is bounded from below by  the relative entropy of entanglement 
\cite{VPRK1997,PlenioVedral1998}.
We also find that the CLOCC paradigm allows to define thermodynamical cost of 
erasure of entanglement. The cost is also bounded from below by 
the relative entropy of entanglement.
We also analyze the deficit for multi-party pure states such as 
the Aharonov state, GHZ state and W state. We obtain that, according to the deficit, the Aharonov state 
exhibits the greatest quantum correlations, while GHZ state - the least. 
In fact, the Aharonov state can be said to be quasi-non-localisable, in the sense that in large
dimension, the fraction of information which can be localised goes to zero.
We show that in the finite regime (i.e where Alice and Bob deal with a single copy of a state), 
any entangled state is {\it informationally nonlocal} i.e. it has nonzero deficit. Moreover, 
we provide states which exhibit informational nonlocality even though they
are separable and have an eigenbasis of distinguishable states - call it {\it non-locality 
without
entanglement but with distinguishability} (on the level of ensembles, it has its counterpart in \cite{Bennett-nlwe}).
We also provide many other interesting results, including the impossibility of  catalysis 
with local pure states, and non-existence of states whose entire informational 
contents is non-localisable.

The paper is organized as follows.
After the Introduction, in section \ref{sec:information} an operational
meaning of information is briefly recalled in terms
of transition rates and basic laws of thermodynamics.
In section \ref{sec:restricted}, the idea of information as a resource in
distant labs paradigm is presented. Here the central notion
of the present formalism i.e. {\it quantum information deficit} is defined.
In section \ref{sec:deficit-definition} the various aspects of information
deficit and its dual notion {\it localisable information}
are discussed and an interpretation of the deficit
in the context of quantum nonlocality is provided.
Section \ref{sec:ipb} presents deficit as entropy
production needed to reach set of \pclcor\ states. The concept is then
generalized
to arbitrary set, including set of separable states, and cost of erasure
of entanglement is defined.
Section \ref{subsec:main} provides upper and lower bounds for deficit in
terms
of relative entropy distance and upper bound for entanglement erasure cost.

We next turn to exploring new phenomena which can be discovered using our
methods.
In section \ref{sec:basic}, main implications of the results of previous
section
are provided including the key conclusion that
any entangled state is {\it informationally nonlocal}
in a well defined, natural sense. We also prove the existence of separable
states which
have a locally distinguishable eigenbasis, yet contain nonlocalisable
information.
Section \ref{sec:multi} is devoted to the generalization
to the multipartite case. Some of these results were briefly noted in
\cite{OHHH2001}. Here information deficit is calculated
and the asymptotic behavior is analyzed
for special examples of pure multipartite states:
GHZ state, W-state, and the Aharonov state. We find that the Aharonov state
can be considered to be the most non-local.
Section \ref{sec:Bell} contains exhaustive analysis of
Bell states. In section \ref{sec:noboundinfo} we prove that
(as a opposed to - pure nondistillable
entanglement i.e.. the bound entanglement phenomenon) pure unlocalisable
information does not exists.
Section \ref{sec:super} includes analysis of the
proportions of quantum and classical
correlations in quantum states, addressing
the question: can the first component exceed the second?
In section \ref{sec:zero-one-way} zero-way and one-way subclasses
of informational deficit are presented. It is
shown that in asymptotic version one way
deficit is nonzero for separable (disentangled) states,
stressing that quantum correlations is more than
quantum entanglement.
Section \ref{sec:discord} discusses relation of our measure to other
measures of quantumness of correlations i.e.. {\it one-way and
two-way quantum discord} is discussed.
Section \ref{sec:classical} contains discussion of the result in
the context of classical correlations measure
introduced by other authors including the
Henderson-Vedral measure.
Discussion of complementarity between information quantities
in distributed quantum systems
is provided in section \ref{sec:compl}
The paper closes with general discussion of the results
and a list of open questions in section \ref{sec:conclusions}.

\section{Information: an operational meaning}
\label{sec:information}

Before turning to the case of parties who are in distant labs, it will
prove worthwhile to discuss the notion of information from a more general
perspective.  Although we often talk about information as an abstract
concept, here, we use it as a term of art which refers to a specific function
\be
I(\varrho)=\log_2d-S(\varrho) \s 
\label{eq:infofunc}
\ee
where 
$S(\varrho)=-\tr\varrho\log\varrho$ is the von Neumann entropy
of $\varrho$ acting on a Hilbert space of dimension $d$.  We will usually work
with qubits, in which case $\log d=N$ an integer. As we will see in next section, 
so defined information function has operational meaning: it is number of 
pure qubits one can draw from many copies of the state.  

Let us now shortly discuss the information function (\ref{eq:infofunc}) in the context of more common
Shannon picture. 
In the latter approach a source produces a large amount of information if
it has large entropy. Thus information can be associated with entropy.
This is because the receiver is being informed only if he is
``surprised''. In such an approach the information has a subjective meaning:
something which is known by the sender, but is not known by receiver.
The receiver treats the message as the  information,
if she didn't know it. However, one can also consider an objective picture; 
a system represents information if it is in a pure
state (zero entropy).  We {\it know} what state it is in.
The state is itself the information.  

We obtain a dual picture, where two kinds of information are dual. Shannon's entropy
represents the information one {\it can get to know} about the system,
while the information of (\ref{eq:infofunc}) represents the information one {\it knows }
about the system. Together they add up to a constant, which characterizes the system only (not 
its particular state)
\be
I(total)=\log d = S(\rho) + I(\rho)
\ee

Note that the "objective" picture is more natural in the context of
thermodynamics.  There, a heat bath is highly entropic,
and we are ignorant of exactly what state it is in.  
On the other hand, it is known that using pure states, one 
can draw work from a single heat bath
using a Szilard heat engine \cite{szilard} 
The pure state represents information needed to order the energy of the heat bath.
Knowing which side of a box the molecules of gas are in,
allows one to draw work by having the molecules push out a piston. 
High entropy of the gas, implies ignorance of the molecule positions, 
and an inability to draw work from the system. In general from a single heat bath
of temperature $T$ by use of a system in state $\rho$, 
one can draw amount of work (cf. \cite{Levitin93})
\be
W=kT I
\ee
The process does not violate the second law because the information is depleted
as entropy from the heat bath accumulates in the engine, and one cannot
run a perpetual mobil.  Thus a quantum system in a nonmaximally mixed state 
can be thought of as a type of fuel or resource. 
In fact, originally, our motivation for considering the function \ref{eq:infofunc} in \cite{OHHH2001}
was to understand entanglement in a thermodynamical context.

\subsection{Information and transition rates}
\label{subsec:info-transition} 

In \cite{nlocc,uniqueinfo} it was shown that 
the function $I$ has operational meaning in the asymptotic regime of 
many identical copies. 
It gives the number of pure states that one can obtain from a state $\varrho$
under a certain class of operations we call Noisy Operations (NO):
operations that consist of (i) unitary transformations (ii) 
partial trace and (iii) adding ancillas in maximally mixed state.
First, one can show that it is the unique function (up to constants) 
that is not increasing under the class NO.
One then shows that $I$ determines the 
optimal rate of transitions between states under NO. 
Let us now discuss two special cases. 

First,  given $n$ copies of state  $\varrho$ 
one can obtain $nI(\varrho)$ qubits in a pure state. This is done essentially 
by  quantum data compression \cite{Schumacher1995} (c.f. \cite{PetzMosonyi,HiaiPetz91}).
In data compression, one keeps the signal, and discards the qubits which are in
the pure state.  Here we do the opposite.  We discard the ``signal'', treating
it as noise, and keep instead the redundancies (that are in pure state).
Thus we obtain pure states. 
This is essentially like cooling
\cite{VaziraniS1998}. The protocol does not require 
using noisy ancillas (e.g. maximally mixed states). 

A second protocol of interest is that one can take $n(N-S(\varrho))$ 
pure qubits, and produces $n$ copies of $\varrho$. The protocol, 
described in \cite{uniqueinfo}, takes pure states and dilutes them
using ancillas in the
maximally mixed state (noise). 
Existence of such dual protocols is similar to entanglement 
concentration and dilution \cite{BBPS1996}. 
And, similarly as in \cite{popescu-rohrlich,DonaldHR2001}, this  can be used to 
prove that there is unique function that does not increase under NO class of operations.

Note that for $K$ pure qubits, information is equal to $K$.
For the maximally mixed state $I=0$. 
As mentioned, $I$ is monotonically decreasing under partial trace, and adding ancilla
in maximally mixed state. It is of course constant under unitary 
operations.
The property that makes it a unique measure of information 
in the asymptotic regime is {\it asymptotic continuity} (see 
\cite{Vidal-mon2000,Michal2001,DonaldHR2001})
which means that if two states  are close to each other, then 
so is their informations per qubit. 
It is important to remember that $I$ is not expansible, i.e. 
if we embed the state into larger Hilbert space, then it changes (because 
the number of qubits increases).  
The reason is obvious even within the classical 
framework: if there are two possible states of the system, 
knowledge of the state represents less information than 
knowledge of the state in the case of, say, three  possible configurations. 
It is in contrast with 
entanglement theory where a pure state of Schmidt rank two means 
always the same thing, independently of how large the system is. 
Also entropy of the state depends only on nonzero eigenvalues:
e.g. the entropy of a pure state is zero,  independently  
of how large the system is.  
However in the present case, the Hilbert space and its dimension is 
important element of our considerations.

\subsection{Information in context of "closed operations"}
\label{subsec:CO}
In previous section we have argued that information function 
gives transition rates from mixed state to pure and backwards, and that 
it gives uniqueness of information in the context of NO. For the rest of this paper 
we will not treat additional mixed state as free resource. Thus let us now discuss 
the meaning of information in the context of a class that is compatible 
with the class of operations which we will use in the case of distributed systems 
further in this paper. Namely, we can consider {\it closed operations} (CO).
They are arbitrary compositions of the following two basic operations:
\bei
\item[(i)] unitary transformations
\item[(ii)] dephasing $\rho \to \sum_iP_i \rho P_i$ 
where $\sum_i P_i=I$, and $P_i$ are projectors not necessarily of rank one.
\eei
We call the class closed, though it is not actually fully closed. 
The information cannot go in, but can go out (via dephasing).
The name closed is motivated by the fact that the number of qubits is the same, 
and the qubits cannot be exchanged between the system of interest and environment. 
The only allowed contact with environment is decoherence caused by  operation (ii). 
In next section we will introduce a "closed" paradigm to distant labs scenario,
by use of which we will define quantum deficit.

Now, let us ask what about drawing pure qubits out of given state by the present class of operations.
The operations do not change the size of the system, 
so that when we start e.g. with many copies of state $\rho$ we cannot end up with 
a smaller system in almost pure state.
However this is not a big problem, Imagine for a while that in addition we can apply partial trace (which 
is not allowed in CO). Then the process of drawing pure qubits can be divided into two stages:
1) some CO operations aiming to concentrate the pure part into some number of qubits, and  2) 
partial trace of the remaining qubits.

Since we do not allow for partial trace, one can simply stop before tracing out.The obtained state 
will have a form of (approximate) product of qubits in pure state and the rest of the system 
- some garbage. Thus the process of dividing system into pure part and garbage we can treat as extraction pure 
qubits. 

Now, let us ask how many pure qubits can be drawn from  a state 
by closed operations in the above sense? Actually, the process of drawing qubits by NO
didn't use maximally mixed states. It was just a unitary operation, plus partial trace. 
Thus we can apply this operation (unitaries are allowed in CO) and get again $I$ 
qubits per input states. Thus also within "closed picture" information has the same 
interpretation of maximal amount of pure qubits that can be obtained from a state per input copy,
by closed operations.

\section{Restricting the class of operations in distant labs paradigm: CLOCC  and the information deficit}
\label{sec:restricted}


In the preceding section, we discussed the notion of information from the perspective
of being able to reversibly distill pure states from a given state $\varrho$.  Now,
one can ask about how things change when the allowable class of operations one can perform
are somehow restricted.  This is a rather general question, but since here we are interested
in understanding entanglement and non-locality, we will examine the restricted 
class of operations which occurs when various parties hold some joint state, but are in
distant labs.  One then imagines that Alice and Bob wish to distill as many locally pure
states as possible i.e. product pure states such 
as $\ket{0^{\otimes m_A}}_A\otimes\ket{0^{\otimes m_B}}_B$.  
The amount of local information which is distillable, we call $\Il$.

In the ordinary approach to the distant labs paradigm, one imagines
that two parties (Alice and Bob) are in distant labs and can only perform local operations
and classical communication (LOCC). However, as we noted, this class of operations
is not suitable 
to deal with the questions of concentration of information to local 
form.  That is because under LOCC, one does not count the information that gets added
to the systems through ancillas, measuring devices etc.    
We thus have to state  the paradigm more precisely.
Since we are interested  in  {\it local information},
we must treat it as a resource, assuming it cannot be created,
but only manipulated. Once we have a compound state, the task is 
to localize the information  by using classical channel between Alice and Bob

The new paradigm was introduced in Ref. \cite{OHHH2001} where one essentially looks
at a closed system as one does in thermodynamics when calculating changes in entropy.
One imagines that Alice and
Bob are in some closed box, and don't allow them to import additional quantum states,
except for ones which we specifically keep track of and account for.

In defining a class of operations, the crucial point is that here, unlike in usual
LOCC (local operations and classical communication) schemes, one must explicitly account for
all entropy transferred to measuring devices or ancillas.
So in defining the  class of allowable operations one must ensure that
no information loss is being hidden when operations
are being carried out. Moreover the operations should be 
general enough to represent faithfully the ultimate possibilities 
of Alice and Bob to concentrate information. In other 
words we wouldn't like to introduce any limitation apart from 
two basic ones: (i) there is classical channel between Alice and Bob
(ii) local information  is a resource (cannot be increased).

We consider a state $\rab$ acting on Hilbert space 
$\hcal_{AB}=\hcal_A\ot\hcal_B$.
Let us first define the elementary allowable elements
of Closed LOCC operations (CLOCC). 

\begin{definition}
By CLOCC operations on bipartite system of $n_{AB}$ qubits we mean all 
operations that can be composed out of 
\bei
\item[(i)] local unitary transformations
\item [(ii)] sending subsystems down a completely decohering 
(dephasing) channel
\label{item:cl-channel}
\eei
\end{definition}

The latter channel is of the form
\be
\varrho_{in}\to \varrho_{out}=\sum_iP_i\varrho P_i
\ee
where $P_i$ are one-dimensional projectors. 
For a qubit system, it acts as 
\be
\varrho_{in}=\left[\bea{cc}
\varrho_{11} & \varrho_{12} \cr
\varrho_{21} & \varrho_{22} \cr
\eea\right]
\to 
\varrho_{out}=\left[\bea{cc}
\varrho_{11} & 0 \\
0 & \varrho_{22}\\
\eea\right].
\ee
It is understood, that $\varrho_{in}$ is at the sender's site, while $\varrho_{out}$
is at the receiver's site. The operation (ii) accounts for both local measurements
and sending the results down a classical channel.  It can be 
disassembled into two parts: a) local dephasing  
(at, say, sender site) and b) sending a qubit intact 
(through a noiseless quantum channel) to receiver
Thus suppose that  Alice and Bob share a  state 
$\varrho_{AB}\equiv \varrho_{A'A''B}$, and Alice decided to 
send subsystem $A''$ to Bob, down the dephasing channel.
The following action will have the same effect. Alice dephases locally the 
subsystem  $A''$
\be
\varrho_{A'A''B} \to \sum_i P_i^{A''} \otimes I_{A'B} 
\varrho_{A'A''B}
P_i^{A''} \otimes I_{A'B}
\ee
The state is now of the form 
\be
\varrho_{A'A''B}^{out}= \sum_ip_i P_i^{A''} \otimes \varrho^{A'B}_i
\ee
Thus the part $A''$ is classically correlated with the rest of 
the system (it is stronger than to say that the state is separable 
with respect to $A'':A'B$). Now Alice sends the system $A''$ 
to Bob through an ideal channel. Thus the final state differs 
from the state $\varrho_{A'A''B}^{out}$ only in that the 
system $A''$  is at Bob site.  It follows that the operation 
\ref{item:cl-channel} can be replaced by the following two operations
\bee
\item[(iia)] Local dephasing
\item[(iib)] Sending completely dephased subsystem. 
\eee
Note  that operations  (i) and (iib) are reversible. Only  the
operation (iia) 
can in general, be  irreversible. Actually it is irreversible, 
if only it {\it changes} the state, i.e. in all nontrivial cases. 
Note also that the operations do not change the dimension of 
the total Hilbert space, or, equivalently, the number 
of qubits of the total system, even though the particular qubits can
be reallocated; for example at the end all qubits can be at Alice's 
site.  

Let us finally ,note that  it may happen that after the protocol,
one of the parties will be left without particles at all,
as all have been sent to the other ones. It is only the total number of 
particles which is conserved. 

\subsection{Comparison with other class of operations}
For the purpose of the present paper, we will use solely CLOCC operations. 
Yet, since in some of our papers a different class of operations was employed, we 
will describe the other class and compare it with CLOCC. 

Let us first present the other (likely equivalent) class of operations, called Noisy LOCC (NLOCC). 
The relation between NLOCC and CLOCC will be similar to the relations between NO and CO:
the elementary operations will be the same as in CLOCC, plus tracing out local systems and adding maximally 
mixed ancillas.

\begin{definition}
By NLOCC operations  on bipartite system of $n_{AB}$ qubits we mean all 
operations that can be composed out of 
\bei
\item[(i)] local unitary transformations
\item [(ii)] sending subsystem down  completely decohering 
(dephasing) channel
\item[(iii)] adding ancilla in maximally mixed state
\item[(iv)] discarding local subsystem
\eei
\end{definition}
As in CLOCC we can decompose $(ii)$ into $(iia)$ and  $(iib)$.  
CLOCC operations is more basic than NLOCC. Namely, the latter can be treated as CLOCC 
with additional resource: unlimited supply of maximally mixed states (which 
have zero informational contents).
Indeed, similarly as in section \ref{subsec:CO} one can argue 
that the operation of local partial  trace is not essential. 

\section{Localisable information and information deficit}
\label{sec:deficit-definition}
In this section we  define the central quantity information deficit. To this end 
we will first introduce notion localisable information. We will first deal with 
single copy case, and define basic quantities on this level. 
Then we will discuss the asymptotic regime, which will require regularization 
of the quantities. 

\begin{definition}
The {\it localisable information} $\Il(\rab)$ of a state $\rab$ 
on Hilbert space $C^{d_A}\ot C^{d_B}$ is the maximal amount
of local information that can be   obtained by CLOCC operations. More formally:
\be
\Il(\rab)=\sup_{\Lambda\in CLOCC} (I(\vr'_A)+ I(\vr'_B))
\ee
where $\vr'_{AB}=\Lambda(\vr_{AB})$, $I$ is information function $I(\vr'_X)=N'_X- S(\vr'_X)$;
$N'_A=\log d'_A,N'_B=\log d'_B$ are  number of qubits of subsystems of 
the output state. When one of numbers of qubits is zero (null subsystem) we apply 
the convention that information is zero. 
\end{definition}
Alternatively, we have formula:
\be
\Il(\rab)= N - \inf_{\Lambda\in CLOCC)} (S(\vr'_A)+S(\vr'_B))
\ee
where $N$ is total number of qubits. Again, if it happens that all 
particles are with one party (i.e. the output dimension is equal to one)
so that the subsystem of the other party is null, 
then we apply convention that the entropy of such subsystem is zero.  
Further state on system with one subsystem null we will call null-subsystem 
states.

It is important here, that "to  obtain local information" does not mean as usual 
getting some outcomes of local measurements. Rather it means, to apply such operation, 
after which, information, as a function of states of subsystems  will be maximal. 
Thus, we only deal with {\it state changes}, and calculate 
some function (information function) on states.

Actually is not localisable information which will be the most important quantity. 
Rather, the central quantity  is a closely connected  one, which we call {\it quantum 
information deficit} (in short quantum deficit). It is defined as  a difference between  
the information that 
can be localized by means of CLOCC  operations, and  total information of the state. 
\begin{definition}
The quantum deficit $\D(\rab)$ of a state $\rab$ 
on Hilbert space $C^{d_A}\ot C^{d_B}$ is given by 
\be
\D(\rab)=I(\rab)-\Il(\rab)
\ee
\end{definition}
Using definition of localisable information $\Il$, we get alternative formula for 
quantum deficit
\begin{equation}
\label{twowaydel}
\D=\inf_{\Lambda\in CLOCC}(S(\varrho'_A) + S(\varrho'_B)) - S(\varrho_{AB}).
\end{equation} 
where $\vr'_{AB}=\Lambda(\vr_{AB})$. 

It is important to notice that both quantities  are functions not 
only of a state but also the dimension of the Hilbert space. This is because CLOCC operations 
are defined for a fixed Hilbert space. 
That $\Il$ depends on 
dimension of the Hilbert space is even more obvious, because the latter 
is explicitly written in the formula. However, in the formula for deficit 
as written in (\ref{twowaydel}), the dimension does not appear explicitly,
so it could happen that there is no dependence on dimension.
Actually, it is rather important that $\D$ does not actually depend on dimension,
i.e. when one locally increase Hilbert space, by e.g. adding a qubit in pure state,
$\D$ should not change. This is because,  as we will see later, 
the deficit will be interpreted as a measure of quantumness of correlations,
which should not change upon adding local ancilla. We will discuss 
this issue later in more detail. In particular in section  \ref{sec:noboundinfo} 
we will show that regularization of deficit {\it does not change} 
upon adding local ancilla in pure state.

\subsection{Interpretation of quantum deficit: measure of "informational nonlocality"}

Nonzero deficit means that Alice and Bob  are not able to localize all the information 
contained within the state. This, however, means that part of information is 
necessarily destroyed in the process of localizing by use of classical 
communication. This part of information cannot survive traveling classical 
channels. It implies, that it must be somehow quantum. In addition, 
this part of information must come from correlations, since information 
that is not in correlations, is already local, and need not be localized. 
We could say that quantum deficit quantifies quantum correlations. 
However, we will see that quantum deficit can be (and often is) nonzero for separable 
states, which can be generated by  local quantum actions and solely classical
communication. 
It is not clear then if we can talk here about quantum correlations, because 
can quantum correlations be created by only classical communication between
the parties?  However, quantum deficit being nonzero indicates that 
there is something quantum in correlations of the state. 
One can say, that these are {\it classical} correlations of {\it quantum} properties. 
We will then propose to interpret quantum deficit as amount of  "quantumness 
of correlations".  

Let us now discuss issue in the context of a notion of nonlocality considered 
by \cite{Bennett-nlwe}. The authors exhibited ensembles of product states which 
are fully  distinguishable if globally accessed, but cannot be perfectly 
distinguished  by distant parties that can communicate only via classical 
channel, Then they called this effect {\it nonlocality without 
entanglement}. The reason for using term "nonlocality" was 
the following: one can do better if the system is accessible as a whole, 
rather than when it is accessible by local operations and classical communication. 

In our case, the situation is similar:  Alice and Bob can do better in distilling 
local information if they have two subsystems at the same place rather than 
shared in distant labs. Thus, we have similar kind of nonlocality, and 
quantum deficit is a measure of such nonlocality, which we can call "informational", 
as it concerns difference in access to informational contents. Thus, any state 
with nonzero deficit will be called informationally nonlocal (or nonlocal,
when the context is obvious).

\subsection{Classical information deficit of quantum states}
It is important to investigate not only "quantumness" of 
compound quantum states, but also the relationships between their "classical" and 
"quantum" parts. To this end consider quantity $\ILO$ - the information that 
is local from the very beginning, i.e. 
\be
\ILO=N-S(\rho_A)-S(\rho_B)
\ee
We will call it local information. 
We can now define an analogous quantity to quantum deficit, on a "lower" level.
\begin{definition}
A classical deficit of a quantum state is a difference between local information 
and the information that can be obtained by CLOCC (i.e. by localisable information)
\be
\Dc=\Il-\ILO
\ee
\end{definition}

This tells us how much more information can be obtained from the state by
exploiting additional correlations in the state $\rho_{AB}$.
We will refer to $\Delta_c$ as the {\it classical deficit}, because the channel
is classical. Also, as we will see later, the quantity can be used 
in the context of quantifying of classical correlations (though it is not immediate,
see \cite{SynakH04-deltacl}).

\subsection{Restricting resources:  zero-way and one-way subclasses}
\label{sec:zero-one-way-defs}

Additional measures of quantumness of correlations which
arise when one restricts the communications between Alice and Bob.

One can define the one-way (Alice to Bob) deficit 
(\(\Delta^\rightarrow \)) and one-way (Bob to Alice) information 
deficit (\(\Delta^\leftarrow \))
by restricting the classical communication to only be in one direction.  Furthermore, one has
also a zero-way deficit (\(\Delta^{\o} \)).  The name {\it zero way} is perhaps confusing.
It refers to the situation where no communication is allowed between Alice and Bob until after
they have completely dephased (or performed measurements) on their systems.  
After they have done
this, they may then communication in order to exploit the (what are now) 
purely classical correlations
in order to localize the information.
These restricted deficits corresponds to 
locally accessible informations \(I_l^\rightarrow\), \(I_l^\leftarrow\)
and \(I_l^{\o}\).

\subsection{Asymptotic regime: Distillation of local information as dual picture 
to entanglement distillation}

In this section we will argue that the idea of localization of information,
though at a first glance exotic, can be recast in terms typical for 
quantum information theory, where of central importance are 
manipulations over resources. Even more, our present formulation will be 
analogous to the scheme which is a basis for  entanglement theory: entanglement 
distillation. We will use the interpretation of the information function 
as the amount of pure qubits one can draw from a state in limit of many copies. 

Instead of singlets our precious resource will be pure local qubit. The aim 
of Alice and Bob is: given many copies of state $\vr_{AB}$  to distill 
the maximal amount of local pure qubits by means of CLOCC operations.
(in entanglement theory, we had LOCC operations, however here 
we need CLOCC, otherwise one could add for free states, and the maximal 
distillable amount of pure local qubits would  be infinite).
One way of doing that is the following:
Alice and Bob take state $\vr_{AB}$, 
apply the CLOCC protocol that optimizes formula for localisable information 
i.e. they obtain state $\vr'_{AB}$ 
which has maximal local informations $I'_A$ and $I'_B$.
They apply such protocol to every copy  of state they share. 
As a result they obtain many copies of state $\vr'_{AB}$.
Now, Alice in her lab, can apply protocol of drawing pure qubits out of 
her state ${\vr'}_{A}^{\ot n}$, obtaining $I'_A$ pure qubits. 
The same does Bob. Finally, they possess $I'_A+I'_B$ pure local qubits
which is equal just to localisable information, and actually it is the best they can 
do, when act first on single copies using communication, and only
locally perform collective actions on many copies. 

Alice and Bob could do better, when they act collectively from the very beginning.
In this way we get that the optimal amount of local pure qubits that can be distilled 
by CLOCC is equal to regularization of localisable information:
\be
\Il^\infty=\lim_n{ \Il(\rho^{\ot n})\over n}
\ee
Similarly we can define regularized quantum and classical deficit
\be
\D^\infty=\lim_n {\D(\rho^{\ot n})\over n};\quad \Dc^\infty=\lim_n {\Dc(\rho^{\ot n})\over n}
\ee
Thus we conclude that regularizations of our quantities 
have operational meaning connected with 
amount of pure local qubits which can be distilled out 
of large number of copies of input state by means of different 
resources (global operations, CLOCC, local operations). 
Let us emphasize here, that when Alice and Bob are given single 
copy of state, they usually cannot distill pure qubits.
When, they are given many copies, the ultimate amount of 
distillable pure qubits is described by regularized $\Il$. 
Thus the non-regularized quantity does not  represent
the amount of pure qubits that can be drawn either from
single copy or from many copies. However, since in definition of $\Il$ 
there is information function that has operational asymptotic meaning, 
then $\Il$ also {\it has} some {\it asymptotic interpretation}, representing  
the amount of pure local qubits that can be drawn  when at the stage of communication, 
Alice and Bob operate on single copies, and only after that stage operate collectively.

In entanglement theory, there is similar situation with entanglement of formation
and entanglement cost. The first is not the ultimate cost of producing a state 
out of singlets, though it already contains "some asymptotics" in definition - 
the von Neumann entropy, which is asymptotic cost of producing pure states 
out of singlet. The ultimate cost of producing states out 
of singlets is regularization of entanglement of formation. 

There is however some difference. 
Namely, even  $\Il$ itself, without regularization, has operational 
meaning. Indeed, it is proportional to the amount of work one can draw
from a single copy of the state in presence of local heat baths. 
To draw optimal number pure qubits, one needs many copies. In one copy case, 
drawing pure qubits is highly non-optimal. 
However to draw optimal amount of work by use of state and single heat bath, 
one copy is enough. roughly speaking, in former case, law of large numbers 
(in other words, ergodicity) 
comes from  many copies, while in the latter - ergodicity 
is "supplied" by  heat bath.

Finally, one can also consider amount of local information that can be distilled 
by means of one-way classical communication. It is equal to regularized 
one-way quantum deficit $\D^\to$. In similar vain we can consider regularizations 
of other quantities based on restricted resources, such as 
$\Delta_c^\to$, $\Delta^\o$, $\Delta_{cl}^\o$ etc. 
Again, all those regularizations have operational meaning.

\subsection{Additional local resources}
One of basic features of the paradigm is that adding local ancillas 
is not for free. The reason is that otherwise, all the quantities 
would become trivial. However there are two kinds of local resources 
that still can be taken into account. 

First of all, we can allow adding for free local ancillas in maximally mixed state.
Thus given a state $\vr_{AB}$ we can ask, what about the quantities 
of interest for the state $\vr_{AB}\ot{I_{A'}\over d}$. 
Note here that this would mean that $\Il^\infty$ does not change, 
if we use NLOCC class instead of CLOCC.  Indeed, as have already mentioned, 
the only difference, between two classes for the problem of distillation 
of local information, may appear when  adding local maximal noise could help. 
In general, upon adding such local noise, localisable information could only go up. 
However it is more likely, that it will not change. 
In fact, Devetak has shown \cite{igor-deficit} that  one-way deficit 
does not change upon adding noise. We were not able to show the same in the case 
of two-way communication, though we believe it is also the case. 

Second possibility is {\it borrowing} local pure qubits. 
This would be most welcome, as it would mean 
that the deficit does not depend on dimension of the Hilbert space as discussed in 
the introduction of  section \ref{sec:deficit-definition}. We actually show that 
it is the case for regularized deficit in Section \ref{sec:noboundinfo}.
For one-way case it is shown also in asymptotic regime in \cite{igor-deficit}.

There is more general possibility: borrowing local ancilla in any mixed state. 
However in asymptotic limit, this is actually equivalent to borrowing noise 
and pure qubits, as in that regime any state can be reversibly composed out 
of noise and pure qubits \cite{nlocc,uniqueinfo}.

\subsection{An example: pure states}
\label{sec:pure}

As we have mentioned, in our definition of quantum correlations, we do not 
speak about entanglement at all. We do not work in the established paradigm of optimal rate of 
transformation to or from maximally entangled states \cite{BDSW1996}.  
We consider distillation of pure product states.
Thus, it was perhaps surprising to find \cite{OHHH2001} that 
for pure states, this definition of quantumness of correlations is just equal to the unique asymptotic 
entanglement for pure states  \cite{BBPS1996,BDSW1996}. 

We shall now see that by taking
as an example 
the singlet
\beq
\ket\singlet={1\over \sqrt 2}(\ket{00}-\ket{11}).
\label{eq:singlet}
\eeq
It is a $2$ qubit state of zero entropy, so it's informational content, as given
by Eq. (\ref{eq:infofunc}) is $I=2$.  We will now see that $\Il=1$.  Clearly, without
communicating, neither party can draw any information from the state, since locally,
the state is maximally mixed.  It turns out that
the best protocol is for Alice to send her qubit down the dephasing channel.
After she has done this, Bob will hold the classically correlated state
\beq
\rho_{CC}={1\over 2}(\braket{00}+\braket{11})
\eeq
from which one can extract $1$ bit of information by performing a cnot gate 
to extract one pure state $\ket 0$.
We thus have that $\D=1$.
One can actually view this process in terms of measurements and classical communication, 
as long as we keep track of the measuring
device.  Alice performs a measurement
on the state to find out if she has a $\ket 0$ or $\ket 1$.  She then tells Bob the result.  Bob
now holds a known state, without having to perform any measurement.  Alice on the other hand, had
to perform a measurement to learn her state.  The informational cost of the measurement is $1$ bit since
a measuring apparatus is initially in a pure state, and must have two possible outcomes.
After the measurement, the measuring device needs to be reset. 
The classical state correlated state $\rho_{CC}$, if held between two parties has $\D=0$.
That the process is optimal for singlet state, is obvious, as this is actually 
the only thing which Alice and Bob can do given a single copy. 
However it is highly nontrivial to show that the regularization of $\Il$ 
is still the same. The optimality of this protocol also 
in many copy case was shown in \cite{nlocc}. It also follows from the general theorem 
we give in this paper, which connects deficit with relative entropy 
distance from some set of states.

In general, it is not hard to see that for an arbitrary pure state, the same protocol can 
be used with Alice first performing local compression on her state.
For any pure state \(\left|\psi\right\rangle_{AB}\), the two-way \(\Delta\) is 
given by \cite{OHHH2001,nlocc}
\[\Delta(\left|\psi\right\rangle) = S(\tr_A(\left|\psi\right\rangle \left\langle \psi\right|)).\]
Thus for pure bipartite states, quantum deficit is equal to entanglement. 
It is quite interesting that we have obtained 
entanglement by {\it destroying} entanglement.

\section{Deficit as production of entropy necessary to reach \pclcor\ states}
\label{sec:ipb}

In this section we will show that the quantum deficit can 
be interpreted as the amount of entropy one has to produce 
in the process of transforming a given state into a so-called \pclcor\cite{OHHH2001}.
This expression of deficit makes it possible to define 
entropy production connected with a given subset of states. 
For example we can then speak about the entropy 
production needed to reach the set of separable states. 
In this way our paradigm provides a consistent definition 
of {\it thermodynamical cost of erasure of entanglement}, 
while the original deficit can be called {\it thermodynamical 
cost of erasing quantum correlations}.

\subsection{Important classes of states}

Let us first define sets of states which are important for our analysis.  Notice that in place 
of a simple dichotomy between separable and entangled states \cite{Werner1989}, one can have 
a whole hierarchy of levels of quantumness \cite{habil}. Already Werner 
recognized \cite{Werner1989}, that within entangled states  there might be ones 
that do not violate Bell's inequalities (cf. \cite{Popescu1995,HHH1997-distill}).
One may also go in {\it converse} direction, and within separable states 
find a subclass which is most classical, and wider classes 
which are still somehow classical, though in some sense to a lesser degree 
(cf. \cite{Zurek-discord}). 

First, let us consider a set if states which we choose to call {\it properly \clcor} 
or shortly \clcor. These are states of the form
\be
\vr=\sum_{ij} p_{ij} |i\>\<i|\ot |j\>\<j|
\label{eq:classical}
\ee
where $\{|i\>\}$ and $\{|j\>\}$are local bases. 
Thus any such a state is classical joint probability distribution 
naturally embedded into a quantum state. Note that the set of \clcor\ states 
is invariant under local unitary operations. The states are diagonal in 
a special product basis, which can be called {\it biproduct basis}.

Now let us define the set of states of our central interest. We will call 
them {\it \pclcor\ states } and denote by $\pc$. 
These are the states that can be {\it reversibly} transformed 
into \clcor\ ones by CLOCC. "Reversibly"  means that 
no entropy is produced during the protocol. This implies that no dephasing is needed in 
transformations: Alice and Bob use only unitaries and sending such subsystems 
such that dephasing does not change the total state. Thus 
they can send only such subsystems $X$, that are in the following state 
with the rest $R$: $\rho_{XR}= \sum_ip_i |i\>_X\<i|\otimes \rho^R_i$.
The states that can be in such a way transformed into classical ones can be also 
described  as  the set of states which Alice and Bob
can create under the allowed class of operations (CLOCC) out of classical states.  
The eigenbasis of these states was
called an {\it Implementable Product Basis} or IPB in \cite{nlocc}, since it is 
the eigenbasis that Alice and Bob are able to dephase in. 

Let us note that one can have an intermediate class, {\it one-way classically correlated}
which are of the form 
\be
\vr=\sum_{ii} p_{ii} |i\>\<i|\ot \vr_i
\label{eq:one-way-PC}
\ee
These are states which can be produced out of \clcor\ states 
by one-way reversible CLOCC. They are diagonal in 
basis which is of the form $\{|i\> |\psi^{(i)}_k\>\}$
where $\{|\psi^{(i)}_k\>\}$ are bases themselves. 

The above sets are {\it proper} subsets of separable states, 
and all the inclusions between them are proper too. 

\subsection{Formula for quantum deficit in terms of \pclcor\ states}
\label{subsec:main-theorems}
Any protocol of attaining the information deficit looks as follows:
Alice chooses a subsystem of her system, dephases it, then sends 
it to Bob. Bob then chooses a subsystem from his system (which now includes 
his original system and the system sent by Alice). He dephases his chosen part,
and sends it to Alice. They can send the states using an ideal channel, 
as the sent subsystems are already dephased. Thus sending is here only reallocating 
subsystems, nothing more. Alice and Bob continue such a process as long as they wish. 
When they decide to stop, the final step is $\rho'$ and 
the obtained local information is  equal to $N-S(\rho'_A)-S(\rho'_B)$ 
while the initial total information was $\I=N-S(\rho_{AB})$. 
Thus the deficit obtained in a particular protocol $\pcal$
is $\Delta_\pcal=S(\rho'_A)+S(\rho'_B) -S(\rho_{AB})$.  Alice and Bob 
wish this quantity to be minimal. Suppose then 
that they preformed an optimal protocol, for which indeed  this value 
is minimal. 

There are two  cases: (i) one of subsystems is null (all particles with the other party)
or (ii) both parties have subsystems that are not null. 
Note that in the second case the system must be in product state. 
Suppose it is not. Then, Alice and Bob can dephase state 
in eigenbasis of states of local subsystems. This will not change 
local entropies, but will transform state into \clcor\ one. Then 
Alice can send her part to Bob, so that the information 
contents of the total state will be unchanged. 
However, if only the state was non-product, the total information 
was greater than sum of local informations. 
This means that the protocol was not optimal, so that we have contradiction.

Thus we conclude, that the optimal protocol ends up with either product state
or state of a system,  which one of subsystems is null (all particles either with Bob  
or with Alice).  Even more, when a state is product, one of subsystems 
can be sent to other party, so that the whole system 
is with one party. This is compatible with the philosophy of "localizing"
of information. 

However it turns out that we can divide the total process of localizing of 
information into two stages:
\bei
\item Irreversible stage: transforming input state $\vr$ into some \pclcor\ one $\vr'$ 
\item Reversible stage: localizing information of the state $\vr'$.
\eei
In the first stage Alice and Bob try to produce the least entropy. 
The amount of information that they are able to localize 
is determined by this stage. In second stage, the entropy is not produced, 
and the information is constant. 

We have the following proposition
\begin{proposition}
\label{prop:deficit-cost}
The quantum deficit is of the form
\be
\D=\inf_\pcal(S(\rho')-S(\rho))
\ee
where the infimum is taken over all CLOCC protocols 
that transform initial state $\rho$ into \pclcor\ state $\rho'$.
\end{proposition}

\begin{proof}
The proof actually reduces to noting, 
that \pclcor\ states can be reversibly created from states 
with one null system. Simply, by definition \pclcor\ states 
can be reversibly produced out of \clcor\ states.
The latter, in turn, can be reversibly produced out of 
one-subsystem states. Thus, consider optimal protocol 
for drawing local information.  As we have argued, 
it can end up with a one-subsystem state. Out of the state 
we can reversibly create \clcor\ state which is 
a special case of \pclcor\ states. Conversely, 
suppose that we have a protocol, that ends up with a 
\pclcor\ state. Then one can reversibly transform 
it into one-subsystem state.
\end{proof}

\begin{figure}
\includegraphics[scale=1]{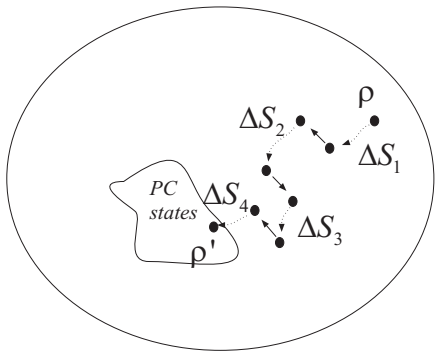}
\caption{CLOCC protocol of concentration of information to local form is a
series of actions aiming to reach the set of \pclcor\ states. 
The continuous lines denote reversible actions: sending dephased 
qubits or local unitary transformations. The  dotted  lines denote dephasings. 
The goal is to make the total entropy increase $\Delta S=\Delta S_1 +\Delta S_2 +\ldots$
minimal. Then the deficit is given by $\D=\Delta S$, because once the state 
is \pclcor, its full information content
 can be localized. 
}
\label{fig:protocol}
\end{figure}

Thus quantum deficit  is equal to {\it minimum entropy production} during making 
state to be \pclcor\ by CLOCC operations. In other words, 
to draw optimal amount of local information from a given state,
one should try to make it \pclcor\ state 
in the most gentle way, i.e. producing the least possible amount of entropy.
Once the state is \pclcor,  the further process of localization of entropy 
is trivial.  The first stage is  illustrated in figure \ref{fig:protocol}.

\subsection{Defining cost of erasing entanglement}

The above formulation of the deficit allows one to generalize the 
idea of thermodynamical cost to other situations. Namely, instead 
of the set of \pclcor\ states one can take any other set 
and ask the same question: how much entropy must be produced,
while reaching this set by use of CLOCC. Thus our concept 
of localizing information allows to ascribe thermodynamical 
costs to other tasks than localizing information. 
With any chosen set we can associate a suitable deficit $\Delta_{Set}$.
An important application of this concept is to take set of separable states. 
Then the associated deficit $\Delta_{sep}$ has interpretation 
of {\it thermodynamical cost of erasing entanglement}. As such it is 
a good candidate for  an entanglement measure. In this paper we will
show that it is bounded from below by relative entropy of entanglement. 
Since set of separable states is a superset of \pclcor\ states,
we have 
\be
\Delta_{sep}\leq \D
\ee
so that the cost of erasing all quantum correlations  is no smaller than 
cost of erasing entanglement. 
For sake of further proofs, let us put here formal definition 
of $\Delta_{sep}$

\begin{definition}
The thermodynamical cost of erasing entanglement $\Delta_{sep}$ is given by 
\be
\Delta_{sep}(\rho)= \inf_\pcal(S(\rho')-S(\rho)
\ee
where infimum runs over all CLOCC protocols $\pcal$ which transform 
initial state $\rho$ into separable output state $\rho'$
\end{definition}

\section{Relations between deficit and relative entropy distance}
\label{subsec:main}
In this section we will  present the proof of the theorem relating deficit to 
in terms of relative entropy distance obtained in \cite{nlocc}. 

\begin{theorem}
\label{thm:main}
The information deficit is bounded from above by the relative entropy distance from 
the set of \pclcor\ states. 
\beq
\D(\rab)\leq \inf_{\sigma\in \pc} S(\rab|\sigma)\equiv \reldist
\eeq
where $S(\vr|\sigma)= \tr \vr \log \vr - \tr \vr \log \sigma$.
\end{theorem}
Let us first prove the proposition
\begin{proposition}
\label{thm:Nipb-Il}
Localisable information and deficit satisfies the following bounds:
\ben
\label{eq:il-H}
&&\Il(\rn)\geq N-\inf_{\bcal\in IPB}H(\rn,\bcal) \\
\label{eq:D-H}
&&\D(\rn)\leq \inf_{\bcal\in IPB}H(\rn,\bcal) - S(\rn)
\een
where $H(\rho,\bcal)$ denotes  the entropy of diagonal entries 
of state $\rho$ in basis $\bcal$
\be
H(\rho,\bcal)=-\sum_ip_i \log p_i,
\ee
with $p_i=\<\psi_i|\rho|\psi_i\>$, with $\psi_i\in\bcal$. 
\end{proposition}
\noindent{\bf Proof.} 
We will exhibit a simple  protocol to achieve a reasonable amount of 
local information. Namely, Alice and Bob choose some implementable 
basis $\bcal$ and dephase a state in such basis. They can do this, as by definition, 
an IPB is a basis in which Alice and Bob can dephase by use of CLOCC. 
The final state has entropy 
\be
S(\rho')=H(\rho,\bcal),
\ee
Alice and Bob can now choose the basis, that will produce the smallest possible 
entropy $H(\rho,\bcal)$. In this way we obtain the following bound for $\D$:
\be
\D\leq \inf_{\bcal\in IPB} H(\rho,\bcal) - S(\rho)
\label{eq:upper-bound}
\ee
This ends the proof of proposition. \blacksquare

Let us now  express this bound in terms of relative entropy 
distance.  This is done by the following lemma. 

\begin{lemma}
\label{Shannon_relent_lemma}
Given a state \(\varrho\), 
\begin{equation}
\label{Shannon_relent}
H(\varrho,\bcal) = \inf_{\sigma \in \scal_\bcal} S(\varrho|\sigma)  +  S(\varrho),
\end{equation}
where \(H(\varrho,{\bcal})\) is the Shannon entropy 
of the probability distribution of the outcomes 
when \(\varrho\) is measured in a given basis 
\(\bcal\) and \(\scal_\bcal\) is the set of all states with eigenbasis \(\bcal\).
\end{lemma}

\noindent{\it Proof.} We have
\ben
&\quad& \inf_{\sigma \in \scal_\bcal} S(\varrho|\sigma) + S(\varrho)\\ \nonumber
&=&   \inf_{\sigma \in \scal_\bcal} (- \tr(\varrho \log_2 
\sigma)) \\ \nonumber
&=& - \tr(\varrho_\bcal \log_2 \varrho_\bcal) + \tr(\varrho_\bcal 
\log_2 \varrho_\bcal) \\  \nonumber
&\quad& +    \inf_{\sigma \in \scal_\bcal}   (- \tr(\varrho_\bcal \log_2 
\sigma)) \\ \nonumber
&=& S(\varrho_\bcal) +  \inf_{\sigma \in \scal_\bcal} S(\varrho_\bcal|\sigma)\\ \nonumber
&=& H(\varrho,\bcal).
\een
Here \(\varrho_\bcal\) is the state \(\varrho\) dephased in the basis \(\bcal\). In the second equality, we
have used the fact that \(\tr(\varrho \log_2 \sigma) = \tr(\varrho_\bcal \log_2 \sigma)\),
because $\sigma$ is diagonal in basis $\bcal$.
In the fourth equality, we have used that \(\varrho_\bcal\) belongs to the set \(\scal_\bcal\) so that 
\(\inf_{\sigma \in \scal_\bcal} S(\varrho_\bcal|\sigma)=0\), and also that 
\(S(\varrho_\bcal) \equiv H(\varrho,\bcal)\). 
This ends the proof of the lemma. \blacksquare

Now combining the lemma with the proposition we obtain the above theorem. 
We have not been able to prove equality, and in subsection \ref{subsec:noncommut} we 
discuss the origin of the difficulties. 

\subsection{Deficit, cost or erasure of entanglement and relative entropy of entanglement}

In the previous section we have reproduced the result of \cite{nlocc} which provided 
upper bound for deficit in terms of relative entropy distance 
from \pclcor\ states. 
In this section we will prove a new result, providing a {\it lower bound 
for the deficit} in terms of an entanglement measure  - the relative entropy of entanglement. 

\begin{theorem}
\label{th:deficit-er}
For any bipartite state $\rho$ the quantum deficit is 
bounded from below by relative entropy of entanglement
\be
\D(\rho)\geq E_r(\rho)
\ee
\end{theorem}

To prove the above theorem it is enough  to show 
that $\Delta_{sep}$ - the cost of erasing entanglement  - 
is lower bounded by $E_r$, which is the contents of 
the next theorem. Indeed, by definition of $\Delta_{sep}$ 
and by the proposition \ref{prop:deficit-cost} 
the deficit is no smaller than $\Delta_{sep}$.

\begin{theorem}
\label{th:cost-er}
For any bipartite state $\rho$ the quantum deficit is 
bounded from below by relative entropy of entanglement
\be
\D_{sep}(\rho)\geq E_r(\rho)
\ee
\end{theorem}

To prove this theorem we will need the following lemma.

\begin{lemma} 
Consider any subset $S$ of states, invariant under product unitary transformations. 
Then relative entropy distance from this set $\ers$ given by 
\be
\ers=\inf_{\sigma\in\scal} S(\rho|\sigma)
\ee
decreases no more than entropy increases under local dephasing, that is 
\be
\ers(\rho)-\ers(\Lambda(\rho)) \leq S(\Lambda(\rho))-S(\rho)
\ee
\label{main}
where $\Lambda$ is local dephasing. 
\end{lemma}

\noindent{\it Proof.}
Note first that local dephasing can be represented as mixture 
of local unitaries:
\be
\Lambda(\rho)=\sum_i p_i U^i_A\otimes I_B\, \rho\, {U_A^i}^\dagger\ot I_B
\ee
Indeed, consider any set of projectors $\{P_j\}_1^k$.
The suitable unitaries are given by 
\be
U(s_1,\ldots,s_k)=\sum_{j=1}^k s_j P_j
\ee
where $s_j=\pm 1$ are chosen at random. Thus $p_i$'s are equal, but this is irrelevant 
for our purpose. 

Now, let us rewrite the inequality (\ref{main}) as follows:
\be
\ers(\rho)+S(\rho)\leq S(\Lambda(\rho))+\ers(\Lambda(\rho))
\ee
Thus we have to prove that function $f(\rho)=\ers(\rho)+S(\rho)$  
is nondecreasing under dephasing. This is somehow parallel result to 
the result of \cite{LPSW1999} where it was proven that the above function 
does not decrease under (global) mixing. The proof is directly inspired 
by \cite{lock-ent}. 

We have 
\ben
&&f(\Lambda(\rho))=\inf_{\sigma\in S} -\tr \Lambda(\rho)\log\sigma= \\ \nonumber
&&=\inf_{\sigma\in S}\sum_ip_i \tr \rho_i \log \sigma\geq \sum_ip_i \inf_{\sigma\in S} 
\tr\rho_i\log 
\sigma=\\ \nonumber
&&=\sum_ip_i \inf_{\sigma\in S} \tr \rho \log \sigma_i =\sum_ip_i \inf_{\sigma\in S} 
\tr \rho \log \sigma\\ \nonumber
&&=f(\rho)
\een
where $\rho_i=U_A^i\otimes I_B \rho {U_A^i}^\dagger\otimes I_B$
and $\sigma_i={U_A^i}^\dagger\otimes I_B \sigma U_A^i\otimes I_B$. 
The inequality comes from properties of infimum, the last but one equality comes from 
the fact that the set $S$ is invariant under product unitary operations.
This ends the proof of the lemma. 
\blacksquare 

{\it Proof of the theorem \ref{th:cost-er}.}
The basic ingredient of the proof is monotonicity of the function 
$f(\rho)=E_r(\rho)+S(\rho)$ under CLOCC. 
(In entanglement theory important functions are the ones 
that cannot increase under suitable class of operations, 
while here we need a function that does not {\it decrease} under 
our class of operations. This once more shows that our approach 
is in a sense dual to the usual entanglement theory.)
As we have already discussed,
any CLOCC operation can be decomposed into basic ones: (i) local unitary 
transformation, (ii) local dephasing and (iii) noiseless sending of dephased
qubits. Of course local unitary operation does not change 
either entropy or $E_r$, so that the function $f$ 
remains constant. The lemma we have just proved tells us that local 
dephasing can only increase the function $f$. Consider now the last 
component - sending dephased qubits. Clearly entropy again does 
not change during such operations. It remains to show that 
$E_r$ does not change under sending dephased qubits.  
Consider the state $\rho_{ABB'}$ with one dephased qubit $B'$ on Bob's site. 
Consider the closest separable state to the state $\sigma_{ABB'}$.
Since relative entropy of entanglement is in particular monotone under dephasings,
we can choose this state to have the qubit $B'$ dephased too. 
Consider then state $\rho_{AA'B}$, where $A'$ qubit is the $B'$ qubit after being
sent by Bob. We now apply the procedure of sending qubit $B'$ 
to the state $\sigma_{ABB'}$ and  obtain a new separable state $\sigma_{AA'B}$.  
By construction we have $S(\rho_{ABB'}|\sigma_{ABB'})=S(\rho_{AA'B}|\sigma_{AA'B})$.
Thus $E_r$ could only go down. However we can repeat the reasoning 
with the qubit sent in converse direction, and conclude that $E_r$ 
does not change.  

In this way we have shown that the function $f$ cannot 
decrease under CLOCC operations. This means, that for any protocol 
that brings initial state $\rho$ 
to a final separable state $\rho'$ we have 
\be
f(\rho')\geq f(\rho)
\ee
However the target state is separable, hence it has $E_r=0$. We obtain
\be
S(\rho')-S(\rho)\geq E_r(\rho)
\ee
which tells us that in any protocol that ends up with separable state,
the increase of entropy is no smaller than relative entropy of entanglement. 
This ends the proof. \blacksquare

\subsection{Connection with bounds obtained via semidefinite programming}
In \cite{SynakHH04} semidefinite programming techniques were used to 
obtain lower bounds on regularized deficit. The following general bound 
was obtained:
\be
\D^\infty(\rho) \geq \sup_\sigma\left(-\log_{2} 
\lambda_{\max}(|\sigma ^{\Gamma}|) -S(\varrho)  
-S(\varrho|\sigma)\right)
\label{eq:sdp-bound} 
\ee
where $\lambda_{\max}$ denotes the greatest eigenvalue, and 
$\Gamma$ is partial transposition of the matrix.  
The value of the bound has been calculated for Werner states and isotropic states. 
It turned out that for those states it is exactly equal to regularized relative entropy 
of entanglement.  This is compatible with the theorem \ref{th:cost-er}. 
It is interesting, what is general relation of the bound (\ref{eq:sdp-bound})
with regularized $E_r$.

\subsection{Discussion of the problem of "noncommuting choice"}
\label{subsec:noncommut}
We have proved that deficit satisfies the following inequality
\be
\reldist\geq \D \geq E_r
\ee
Yet we have not been able to prove that $\D=\reldist$. Let us discuss 
the main obstacles which we encountered. The question is actually as follows:
 Can there be better protocol than
dephasing in optimal IPB basis? The latter protocol has some fundamental feature.
Namely, in the series of the subsequent local dephasings, each dephasing is 
compatible with the previous one in the sense that they commute with 
each other. In other words, each dephasing is in some sense ultimate: it divides 
the total Hilbert space into blocks, so that all subsequent dephasings 
are performed within blocks, in the basis that is compatible with the blocks.
Another way of viewing it is to say that what was sent from Alice to Bob 
or vice versa, will  remain classical, that is diagonal in fixed 
distinguished basis. The main open question is now the following:
Is it enough for Alice and Bob 
to follow this restriction, or whether they should violate this 
rule to draw more information? 

We can formulate this fundamental problem in a more tractable way,
if we look through the proof of the theorem \ref{th:cost-er}, 
and find where the proof fails if instead of separable states 
one takes \pclcor\ states. Almost the entire proof can be carried forward
without alteration, apart from one small item:
the invariance of $\reldist$ under 
sending dephased qubits. $E_r$ 
was invariant mainly because we could choose the closest separable 
state to be also dephased on that qubit. This is because 
set of separable states is closed under local dephasings. 
However the set of \pclcor\ states is not. It does not rule out the possibility 
that indeed the closest \pclcor\ state has the qubit dephased.
However we were not able to prove it or disprove. We will formulate 
here the problem in a formal way:

{\bf Problem.}
{\it Consider bipartite state that can be written in the following form:
\be
\rho_{AB}=p_1 \rho^1_{AB} + p_2 \rho^2_{AB} 
\ee
where $\rho^1_{AB}$ and $\rho^2_{AB}$ are orthogonal on subsystem $A$,
i.e. the reduced states $\rho^i_{A}$ have disjoint support.
Can the closest \pclcor\ state in relative entropy distance 
be written in this form?}




\subsection{Deficit and relative entropy distance for one-way and zero-way
scenarios}

Finally let us note that needed results  can be obtained easily 
for one-way and zero-way scenarios.  
The problem with two-way 
is  that Alice and Bob could draw more information than they obtain by measuring 
in optimal IPB basis. The source of difficulty was that 
in many rounds protocol, Alice  and Bob could make dephasings 
that would not commute with dephasings they made in previous step. 
In the case of one-way there is no such danger, as there is only one round. 
The zero-way situation is simplest. 
The only thing Alice and Bob can do is to dephase the subsystems 
in some bases, and the only problem is to find optimal bases (so that 
they will produce the smallest amount of entropy).
The versions of lemma \ref{Shannon_relent_lemma}  in one-way and zero-way case
can be proven in the same way. Thus in those cases 
the deficits are equal to relative entropy distance to 
the two sets of states - classically correlated states 
and one-way classically correlated states (\ref{eq:one-way-PC}).

\subsection{Multipartite states}
We can define set of \pclcor\ states also in the case of multipartite states.
Then one can formulate version of Theorem \ref{thm:main} in the latter case. 
Since the arguments we have used did not depend on number of parties,
Theorem is then true also in multipartite case. 
Similarly theorems \ref{th:deficit-er} and \ref{th:cost-er} hold in the multipartite 
case.

\section{Basic implications of the theorem (informational nonlocality)}
\label{sec:basic}

The theorems obtained in the previous section allow us to obtain the following results 
for both bipartite as well as multipartite states. 
\bei
\item $\D$ is bounded  no smaller than distillable entanglement $E_D$. 
\be
\D\geq E_D
\ee
Indeed, the latter is bounded from above by relative entropy of entanglement
\cite{Rains2001}.

\item Moreover, Theorem \ref{th:cost-er}
implies that quantum deficit is no smaller than 
coherent information:
\be
\D(\rho)\geq S(\rho_X) - S(\rho)
\ee
where  $X=A,B,C ...$. 
or 
\be
I(\rho)\leq N - S(\rho_X)
\label{eq:Il-coh}
\ee
This is because it was proven that in bipartite case \cite{PlenioVP2000}, 
relative entropy  of entanglement is bounded from below 
by coherent information $S_X-S$. For multipartite states,
one gets it by noting that multipartite relative entropy of entanglement 
is no smaller than the one versus some bipartite cut. 
Then one applies the mentioned bipartite result.



\item {\it Any entangled state 
is informationally nonlocal}, i.e. it has nonzero deficit.

\be
\D(\vr_{entangled})>0
\ee
This follows form the fact that when a state is entangled, then it has nonzero 
relative entropy of entanglement.

Note however that there exist {\it separable} states 
which are informationally nonlocal
\be
\D(\vr_{separable})>0, 
\ee
for some separable states.
We will now discuss 
an example of such state and relate it to 
so called "nonlocality without entanglement".

\item Theorem \ref{th:cost-er} allows for easy proof that for pure bi-partite states 
the deficit is equal to entanglement. Indeed, from the theorem 
we have that deficit is no greater than entanglement. 
On the other hand, a simple protocol of dephasing 
Alice's part in eigenbasis of the state of her subsystem, 
and sending the stuff to Bob gives the amount of information $2 \log d - S(\rho_A)$.
Thus deficit is also no greater than entropy of subsystem. 
However the latter is equal to relative entropy of entanglement 
(this is reflection of the fact that in asymptotic regime 
there is only one measure of entanglement for pure states). 
For multipartite pure states there does not exists 
unique entanglement measure. 
We have the following open question:
{\it For multipartite pure states, is the deficit equal to 
relative entropy of entanglement?}
\be
E_r(\psi)\mathop{=}\limits^{?}\Delta(\psi)
\ee
If so, deficit would be an entanglement measure for all pure states.
And since deficit is an operational quantity, we would 
have {\it operational} interpretation for  relative entropy of 
entanglement for pure states. 

Note here that in general the deficit is not monotone under LOCC, 
and even under CLOCC. In contrast, $I_l$ is monotone under CLOCC. 

\item From the above  reasoning and theorem \ref{th:cost-er} it follows that 
 {\it thermodynamical cost of erasure of entanglement of pure states 
is equal to their entanglement. } (c.f. \cite{OHHH2001,nlocc})
\eei

\subsection{Non-locality without entanglement and with distinguishability}
\label{sec:nlwewd}

One form of non-locality we are familiar with, is entanglement.  Another form of 
non-locality was introduced in \cite{Bennett-nlwe}: so-called {\it non-locality without entanglement}.
There, it was shown that there are ensembles of states, which, although  product, cannot
be distinguished from each other under LOCC with certainty.  Ensembles of  product states can 
have a form of non-locality. Other ensembles were exhibited, which 
were distinguishable, but distinguishing was 
thermodynamically irreversible. 
This can be thought of of as {\it non-locality without
entanglement but with distinguishability}.   All those results were done 
for ensembles. 

Here we report similar kinds of nonlocality for states.
Namely, we will exhibit states which are separable,
and which can be created out of ensembles of distinguishable states but which contain
unlocalizable information such that $\D\neq 0$ (at least for single copies). 
In fact, one can find such states which
have an eigenbasis where each eigenket is perfectly distinguishable.  

An example is the state  given by
\beq
\rho=\frac{1}{4}\braket{00} + \frac{1}{4}\braket{11} + \frac{1}{2}\braket{\singlet}  \s.
\label{eq:mfs}
\eeq
It is a separable state, which can be seen either by construction, or because it has positive partial
transpose which is a sufficient condition for dimension $2\otimes2$.
It's eigenkets $\ket{00},\ket{11},\ket{\singlet}$
are clearly perfectly distinguishable under LOCC, since Alice and Bob just need to measure
in the computation basis and compare results to know which of the three basis state they have.   
Nonetheless, it clearly has non-localisable information. 
To localize all the information, one would need to dephase it
in the basis $\ket{00},\ket{11},\ket{\singlet}$, but this cannot be done under CLOCC, since
one cannot dephase using a projector on $\ket{\singlet}$.
The proof follows from Theorem \ref{thm:Nipb-Il} - we know that the optimal protocol is for
Alice to dephase her side in some basis, and then send the state to Bob.
Indeed, for two qubits, all implementable product bases 
are one-way implementable, i.e. they are 
of the form $\{|i\> |\psi^{(i)}_k\>\}$
where $\{|\psi^{(i)}_k\>\}$ are bases themselves. 
Thus for the one copy case, which we consider here,
the optimal protocol is one-way protocol. Since the state is symmetric, 
then it does not matter which way (from Alice to Bob or vice-versa).

A direct calculation shows that the optimal basis is $\ket{0\pm 1}$ at one of the sites.
This yields $\Il=3/4 \log 3 -1$, while $I=1/2$ giving a value of $\D=.1887$.
There are thus separable states which exhibit non-locality in that all the information cannot
be localized even though all the basis elements of the state are perfectly distinguishable.

\section{Informational nonlocality of multipartite states}
\label{sec:multi}

The approach considered here turns out to be quite valuable in the case of multipartite
states.  One of the reasons for this is that one can not only quantify the quantumness
of correlations along various splittings, as is commonly done, but one can also look
at the total amount of localizable information that a given state possesses if all
parties cooperate.  In other words, in addition to the various {\it vector measures}
defined for a particular splitting of the state eg. $AB|CD$, one also has a {\it scalar
measure} which is defined for the state as a whole.  One can calculate $\D$ for various
bipartite splitting by grouping parties together, or one can calculate $\D$ for the entire
state.  In fact, one can consider all possible groupings, such as $AB|CD|EF$ etc.
This allows one to explore multipartite correlations in more detail, and also allows one
to ascribe a single quantity to a particular state in order to rank various states in terms
of their total quantum correlations.  

By considering a family of states for a number of parties $\N$, one can calculate 
the information deficit per party $\D(\rho_N)/N$. 
and we find that it goes to zero for
the generalized GHZ, and to infinity for the Aharonov state, as $\N$ goes to infinity.
Of the states we consider, we shall thus find that the Greenberger-Horne-Zeilinger (GHZ)
state is the least informationally nonlocal, while the so-called Aharonov state 
is the most informationally nonlocal.

\subsection{Schmidt decomposable states}
\label{subsec_abcdef}

The information deficit for the N party GHZ state
\beq
\ket{\psi_{NGHZ}}=\ket{111...1}+\ket{222...2}+\ldots+\ket{NNN...N}
\eeq
where we depart slightly from convention by taking the dimension of each
parties state to also scale like $N$. This state is thus more entangled than
if one were to give each party a qubit, and we do so in order to fairly compare our 
results with other entangled states.
The deficit for the GHZ was calculated in \cite{OHHH2001} where it was found to be 
$\D(\psi_{NGHZ})=\log N$.  
Essentially, once one party makes a measurement, all the other parties can learn which
state they have without performing a measurement, thus $\Il=(N-1)\log N$, while the total state is of
dimension $N^N$, hence $I=N\log N$.  Therefore 
\beq
\lim_{N\rightarrow\infty} \D(\psi_{NGHZ})/N = 0
\eeq
This is in keeping with the notion that the GHZ is rather fragile, since if only
one of the qubits becomes dephased, the entire state becomes classical.

One can generalize this to any multipartite state which can be written in a Schmidt
basis.  I.e.
\beq
\ket{\psi_{NS}}=\sum_i c_i \prod_{n=1}^N \ket{\phi_{Ni}}
\eeq
In that case, one finds $\D(\psi_{NS})=S(\rho_A)$ where $\rho_A$ is any of the subsystem
entropies (they are all equal).  This follows directly from 
inequality (\ref{eq:Il-coh})
and it holds in the asymptotic regime of many copies.

\subsection{An example of a non-Schmidt decomposable state: The W state in three qubits}
\label{subsec_dur}

A more complicated example is the ``\(W\) state'' \cite{DurVC00-wstate}
\[
|\psi_{W}\>_{ABC} = \frac{1}{\sqrt{3}} (|100\> + |010\> + |001\> )
\]
and we ask the 
question of how much localizable information $\Il$
can be extracted under one-way CLOCC by using it as a shared state.
Since each party only has one qubit, we can use Theorem
\ref{thm:Nipb-Il} to calculate it.   This is because if each party only
holds a single qubit, the optimal protocol will only need one way
communication, and will be equivalent to having one party measure,
and then tell her results to the other parties who will than hold
a pure state between them.

Let Alice measure her part of the state in basis $\{|e_{i}\>\}$ 
and send the result to Bob and Charlie.
After the measurement 
\ben
|\psi_{W}\>_{ABC} \longrightarrow \varrho_{ABC}=
 \sum_{i} p_{i} |e_{i}\>\<e_{i}| \otimes \varrho_{BC}^{i}
\een
Then Alice obtains the ensemble $\{p_{i}, |e_{i}\>\}$. 
Bob and Charlie obtain ensemble $\{p_{i}, \varrho^{i}_{BC}\}$. 
$\varrho^{i}_{BC}$ are of course pure states.
Bob and Charlie know, which of the states $\{\varrho^{i}_{BC}\}$ they have, 
because they have obtained information about the result of the measurement by  Alice.  
Therefore, the total amount of information, that can be extracted from $|\psi_{W}\>_{ABC}$ locally,
by such a protocol, is given by
\ben
 I(\varrho_{A}) + p_{1} I_l(\varrho^{1}_{BC}) +  p_{2} I_l(\varrho^{2}_{BC})
\een
where $\varrho_{A} = \sum_{i} p_{i} |e_{i}\>\<e_{i}|$
 so that
\ben
I(\varrho_{A})  =  1-H(\{p_{i}\})
\een
and where (since \(\varrho_B^i\) are pure)
\ben
 I_l(\varrho^{i}_{BC})  = 2-S(\varrho_{B}^{i}),
\een
 with $\varrho_{B}^{i}$ being the  reduced density  matrix of $\varrho^{i}_{BC}$.
So for an arbitrary von Neumann measurement, 
we have that for the \(W\) state, 
$\Il$
is given by
\begin{eqnarray}
\Il(\psi_W)&=& 3-H(\{p_{i}\})-\sum_{1}^{2} p_{i} S(\varrho_{B}^{i}) \nonumber \\
& = &  3-H(\frac{1+|x|^{2}}{3})  \nonumber \\
  \quad & - & \frac{1+|x|^{2}}{3} H({1\over 2} + \frac{\sqrt{-3|x|^{4}+2|x|^{2}+1}}{2+2
|x|^{2}})  \nonumber \\
\quad & - & \frac{2-|x|^{2}}{3} H({1\over 2} + \frac{\sqrt{4|x|^{2}-3|x|^{4}}}{4-2|x|^{
2}}), \nonumber
\end{eqnarray}
where the measurement is performed in the basis \(\{|e_i\>\}\) given by 
\[
\begin{array}{rcl}
|e_{1}\rangle &=& x |0 \rangle + y| 1 \rangle\\
|e_{2}\rangle &=& y^{*}|0 \rangle - x^{*} | 1 \rangle.
\end{array}
\]
One can check that  for von Neumann measurements, the largest amount of local information extractable is 
\( 1.45026\).
It is achieved for measurement in the basis \(\{|e_i\>\}\),
where either $x^2=1/3$ or $x=2/3$ (see Figure \ref{fig:wstate}). Contrary to naive expectations,
dephasing in the computational basis is the worst choice. 
Also the basis $|\pm\>$ $(x=1)$ is not optimal. It is interesting, 
that optimal bases are not incidental.  Rather these are those bases 
for which probabilities  of transition into $|0\>,|1\>$ states are the same 
as the probabilities of getting those states by Alice measuring  
$W$ state in basis $|0\>,|1\>$.
In the regime of single copies,
this protocol is optimal by Theorem \ref{thm:Nipb-Il},
therefore for the \(W\) state, $\Il=1.45026$.  This is less than the amount of localisable
information for the corresponding GHZ state
$\ket{\psi_{GHZ}}={1\over \sqrt2 }(\ket{000}+\ket{111})$, thus we would argue that the W-state exhibits more
non-local correlations.


\subsection{The Aharonov state and quasi-unlocalisable information}
\label{ssec:aharanov}

We next consider the so called, Aharonov ``diamond'' state.
it is essentially given by anti-symmetrizing N N-dimensional states.  For three 
parties, the unnormalized state is
\beq
\ket{\psi_{3A}}=\ket{012}-\ket{021}+\ket{120}-\ket{102}+\ket{201}-\ket{210}
\eeq
and in general it is
\beq
\ket{\psi_{NA}}=\frac{1}{\sqrt{N!}}\sum_{permutations} \epsilon^{a_1...a_N}\ket{a_1...a_N}
\eeq
where $\epsilon^{a_1...a_N}$ is the permutation symbol (Levi-Civita density).

It has the property that if one party measures their state in any basis, and tells their result
to the rest of the parties, they will then still hold another Aharonov 
state of dimension $N-1$. 
Since this is a pure state of dimension $N^N$, the total amount of information is $\I=N\log{N}$.
On the other hand, under the protocol where the parties take turns measuring, it is easy to
see that after each measurement, the other parties will still be left with a locally maximally
mixed state.  Finally, however, there will be two parties left, and they will share a singlet,
which can be converted into $1$ bit of localized information.

The amount of localisable information is therefore $\Il=1$ regardless of how large $N$ is.
This is optimal by Theorem \ref{thm:Nipb-Il} for single copies.
We thus have that 
$
\D(\psi_{NA})/N = \log{N}-1/N
$ 
which grows logarithmically to infinity with $N$. The amount of localisable information per
dimension goes very fast to zero as $N^{-N}$.  The Aharonov state
can then be thought of as a form of unlocalizable information.  One might wonder if one
can make the localisable information strictly zero, as is the case for entanglement
with bound entangled states.  We will soon show that this is not the case.

\subsection{General pure three qubit states}
\label{ssec:puremulti}

In subsection \ref{subsec_abcdef}, we considered the localisable information of Schmidt decomposable 
states. And in subsection 
\ref{subsec_dur}, we considered the W state, an example of a 
non-Schmidt decomposable state. 

Let us here consider the general three qubit \emph{pure} state, which can be 
written in the form \cite{Acinschmidt,Sudberyschmidt}
\begin{equation}
\label{acin_ghyama}
\left|\psi\right\rangle_{ABC} =
a \left|000\right\rangle + b\left|010\right\rangle + c\left|100\right\rangle
+ d \left| 001\right\rangle + e\left|111\right\rangle,
\end{equation}
where only \(a\) need be complex, while the rest of the coefficients are 
real. Of course we have \(|a|^2 + b^2 + c^2 + d^2 + e^2 =1\). 


We again can use Theorem \ref{thm:Nipb-Il} to obtain the amount of
localisable information.  let us suppose that Alice (A) measures in the basis 
 \be
\begin{array}{rcl}
|e_{1}\rangle &=& x |0 \rangle + y| 1 \rangle\\
|e_{2}\rangle &=& y^{*}|0 \rangle - x^{*} | 1 \rangle,
\end{array}
\label{eq:basis}
\ee
and sends the measurement outcome to Bob (B) and 
Charlie (C).

Depending on the measurement outcome, Bob and Charlie share the state
\[
|\psi_{e_1}\> = \frac{1}{\sqrt{p}}
((x^* a + y^* c)|00\>  + x^* d |01\> + x^* b|10\> + y^* e|11\>)
\]
or 
\[
|\psi_{e_2}\> = \frac{1}{\sqrt{1-p}}
((y a - x c)|00\>  + y d |01\> + y b|10\> - x e|11\>)
\]
corresponding to the outcome \(|e_1\>\) or \(|e_2\>\) at Alice, where
\[p = 
|(x^* a + y^* c)|^2 + |x|^2 d^2 + |x|^2 b^2 + |y|^2 e^2
\]
is the probability that \(|e_1\>\) is obtained by Alice.

For such a protocol, the localisable information amounts to 
\begin{eqnarray}
\label{acin_PV}
\Il &= & \sup_{x,y}
\big[3 - H(p)   
 - p S(\tr_A|\psi_{e_1}\> \<\psi_{e_1}|) \nonumber \\
&& -(1-p) S(\tr_A|\psi_{e_2}\> \<\psi_{e_2}|)\big] 
\end{eqnarray}
where we maximize over \(x\) and \(y\) to obtain the highest localisable 
information.   This is an optimal protocol, and thus we obtain $\Il$.
Let us denote the quantity in square brackets as \(I_l^{xy}\).

\begin{figure}[tbp]
\begin{center}
\epsfig{figure= 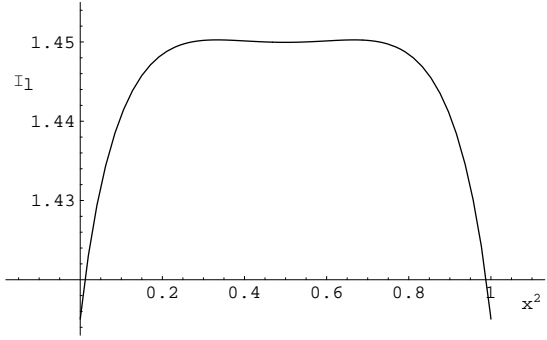,width=0.40\textwidth}
\caption{Plot of $I_l^x$ versus $x^2$ for measurement in basis (\ref{eq:basis}) 
for W-state. The optimal basis for maximizing $I_l^x$ is for 
Alice to dephase (or measure) with $x^2=1/3$ or $2/3$. The basis $|\pm\>$  
($x^2=1/2$) is not optimal. 
}
\end{center}
\label{fig:wstate}
\end{figure}

Let us find the value of the localisable information for 
the case of the \(W\) state \(\left|\psi_W\right\rangle\), using Eq. (\ref{acin_PV}). 
Without loss of generality, we may write \(x = r >0\) and \(y = \exp(i\phi)\sqrt{1-r^2}\). 
We now plot, in Fig. \ref{fig:wstate}, the expression \(I_l^{xy}\) on the \((r,\phi)\)-plane. The supremum can then 
be read off from the figure. 
This supremum will then correspond to the localisable information for the \(W\) state. 
Interestingly, the supremum is attained on two parallel lines on the \((r,\phi)\)-plane. 

\begin{figure}[tbp]
\label{fig_acin}
\begin{center}
\epsfig{figure=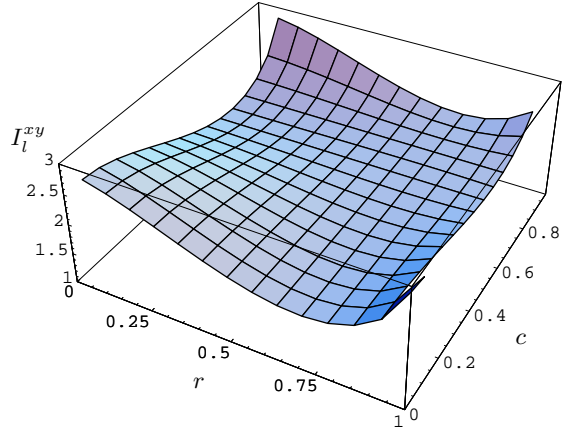,width=0.40\textwidth}
\put(-140,18){\(r\)}
\put(-18,35){\(c\)}
\put(-209,110){\(I_l^{xy}\)}
\caption{Plot of the function  \(I_l^{xy}\) (Eq. (\ref{acin_PV}) 
for the three-qubit state in Eq. (\ref{acin_ghyama}) for the case when \(a=e=0\), \(b=
0.1\),
 on the \((c,r)\)-plane. 
Here \(x=r\), \(y = \sqrt{1 - r^2}\), and \(r>0\). The value of localisable information \(I_l\) 
for a given \(c\) is the supremum of \(I_l^{xy}\) for that value of \(c\). 
} 
\end{center}
\end{figure}

Let us now choose an exemplary one-parameter subclass from the class in Eq. (\ref{acin_ghyama}):
\[
a=e=0, \quad b =0.1.
\]
For this class, we plot the localisable information \(I_l\) using
real values of $x$ and $y$. Taking \(x=r>0\) and \(y = \sqrt{1 - r^2}\), \(I^{xy}_l\) is 
plotted (in Fig. \ref{fig_acin}) as a function of  \(r\) and \(c\). 
For a given \(c\) (which then fixes the state),  the value of \(I_l\) can be 
read from the figure. 

\section{Bell mixtures}
\label{sec:Bell}

The state of eq. (\ref{eq:mfs}) is a particular example of a mixture of Bell states
\[
\begin{array}{rcl}
\displaystyle \left| \phi^{\pm}\right\rangle  & = & {\frac{1}{\sqrt{2}}\left( \left| 00\right\rangle \pm \left| 11\right\rangle \right) },\\
\left| \psi^{\pm}\right\rangle  & = & {\frac{1}{\sqrt{2}}\left( \left| 01\right\rangle \pm \left| 10\right\rangle \right). }
\end{array}
\]

Here, for completeness, we calculate $\D$ for all states of this for - so-called
Bell-diagonal states.  Up to local unitaries, this includes all $2\ot2$ states with local
density matrices that are maximally mixed. 
%
%
%
%
%
%
%
%
Due to Theorem \ref{thm:Nipb-Il}, we only need consider
optimizing
over projection measurements (without adding any
ancilla locally) at one of the parties, say Alice. 
 Consider therefore
the mixture
\begin{equation}
\label{Bellmixture}
\varrho_{Bm} = 
p_1P_{\phi^+} + p_2P_{\phi^-} + p_3P_{\psi^+} + p_4P_{\psi^-}  
\end{equation}
of the four Bell states in \(2 \otimes 2\).

After an arbitrary projection valued (PV) measurement on Alice's side,
projecting in the basis 
\[\left\{\left|\overline{0}\right\rangle = a\left|0\right\rangle
+ b\left|1\right\rangle, \quad 
\left|\overline{1}\right\rangle = \overline{b}\left|0\right\rangle
- \overline{a}\left|1\right\rangle \right\},\]
let 
the global state be projected respectively to 
\beq
P_{\left|\overline{0}\right\rangle} \otimes
\varrho_0, \quad P_{\left|\overline{1}\right\rangle}
\otimes \varrho_1.
\eeq

At this stage, the whole state is essentially on Bob's side. This
is because we allow dephasing as one of our allowed operations. Consequently,
the locally extractable information after this set of operations is the von Neumann entropy of
\[p P_{\left|\overline{0}\right\rangle} \otimes \varrho_0
+ (1-p) P_{\left|\overline{1}\right\rangle} \otimes \varrho_1, 
\]
where \(p\) is the probability of 
Alice obtaining the state \(\left|\overline{0}\right\rangle\).
%
%
The optimization yields 
the value 
\begin{equation}
\label{Bellvalue}
\D=1 + H(p_1 +p_2) - S(\varrho_{Bm})
\end{equation}
where \(p_1\) and \(p_2\) are the two highest  
coefficients of the Bell mixture \(\varrho_{Bm}\).


If we consider only von Neumann measurements (without addition of ancilla) and if Alice and 
Bob are not allowed to make any communication before they perform their measurements, then the 
zero-way information deficit \(\D^{\o}\) for the Bell mixtures (\ref{Bellmixture}) is given by 
\[1 + H(p_{\max}),\]
where 
\[p_{\max} = \frac{1}{2}(1 + |\max\{t_{11},t_{22},t_{33}\}|)\]
with 
\(t_{ii} = \tr(\sigma_i \otimes \sigma_i \varrho_{Bm})\). Note however that 
in this case, we are unable to show whether one can do better by POVMs or whether more copies are 
useful.

Consider however the isotropic $ d \otimes d$ state 
\be
 \varrho_{iso} = \lambda \left| \phi_{max}\right\rangle + (1 - \lambda) \frac{I}{d^2}
\label{eq:iso}
\ee
in \(d\otimes d\),
where \(\phi_{max}\) is the maximally entangled state in \(d\otimes d\) which is invariant under \(U \otimes U^{*}\)
for any unitary \(U\). 
The one-way information deficit \(\D^{\rightarrow}\) (as well as 
\(\D^{\o}\)) is given by
\ben
&&\D^\o=\D^\to=(\lambda+\frac{1-\lambda}{d})\log_{2}(1+\frac{1-\lambda}{d}) \\ \nonumber
&&+(d-1)\frac{1-\lambda}{d}\log_{2}\frac{1-\lambda}{d} - \log_{2}d + S(\vr_{iso})
\een
where 
\ben
&&S(\vr_{iso})= -(\lambda + {1-\lambda\over d})\log (\lambda + {1-\lambda\over d})-\\ 
\nonumber
&&- {d^2-1\over d^2}(1-\lambda) \log {1-\lambda\over d^2} 
\een
For the isotropic state, it is possible to prove, on the same lines as for Bell mixtures,
that POVMs  as well as  more than  one copy cannot help.

\subsection{Asymptotic regime}
\label{subsec:Bell-asymptotic}
For two qubits we easily evaluated deficit, because, one-way and two-way deficits
are equal in  this case, because Alice's first measurement 
leaves no room for other measurements. So the only thing she should
do is to communicate results to Bob, and communication from Bob 
is not needed.  Put it in other words: set of \pclcor\ states is equal to the
one-way classically correlated  states of the form (\ref{eq:one-way-PC}).
Thus it was enough to evaluate only one-way deficit. 
However if we turn to regularization, this equivalence is no
longer valid. This is because, to calculate regularization, 
one needs to evaluate deficit for many copies. Thus the dimension of 
the system is high, and there is  room for many rounds.  
we are not able to regularize two-way deficit. 

Concerning one-way deficit, one can argue that it is 
additive for Bell diagonal states.  Moreover, borrowing qubits  that 
borrowing qubits does not help (It has been independently shown 
that in  general, in one-way case, borrowing pure local qubits 
does not help \cite{igor-deficit}). We will provide the argumentation 
in section \ref{subsec:cl-cor-one-way}.

\section{Purely unlocalisable information doesn't exist}
\label{sec:noboundinfo}

One important aspect of entanglement theory is the existence of bound entangled states. 
These are states which are entangled, in that they require entanglement to create, yet
no entanglement can be drawn from them.  In Section \ref{ssec:aharanov} we saw that in
the multipartite case, there were states which the amount of localisable information per
party went to zero as the number of parties increased.  One can ask whether there is 
a strict analogy to bound entanglement: are there states which have positive $I$, but
which $\Il=0$.  It turns out that the answer is no; the only state which has $\Il=0$ is
the maximally mixed state.  Here we prove this in the following lemma for the case of
two parties. The generalization to many parties is straightforward.
\begin{lemma}
>From any 
 state other than the maximally mixed state  we can draw local information.
\label{lemma1}
\end{lemma}

\begin{proof}
Consider a state  $\varrho \sim C^d \otimes C^d$ such that $ \varrho \neq \varrho _{mmix}=\frac{I}{d^2}$, then  there 
exists an observable  for which the mean value in state $\varrho$ has a different value than
$\varrho _{mmix}$. Every nonlocal  observable can  be decomposed into local operators, so we can always find such 
an observable  of the form $A \otimes B$ for which:

\ben
Tr(A \otimes B)\varrho \neq Tr(A \otimes B)\frac{I}{d^2}
\een
Then
\ben
Tr(A\otimes B)\varrho \neq  \frac{1}{d^2}TrATrB \\
\sum_{ij} p_{ij} \lambda_i \lambda_j \neq \sum_{ij} \frac{1}{d^2} \lambda_i \lambda_j
\label{1}
\een 
Notice that distribution of probability for $\varrho$ in (\ref{1}) is classical. We know that we can obtain a 
nonzero amount of local information from any classical state besides the maximally mixed one.
We can see that we are able to find such local operation, that transforms every state which agrees with the assumptions 
of lemma \ref{lemma1}, into a state from  with we can draw local information.
\end{proof}

There is an open question, whether there exist states, 
for which localisable information is entirely equal to {\it local} information 
content, but which nevertheless are not product. In such case, one wouldn't be able
to draw information from correlations at all. The classical 
deficit $\Dc$ would be zero, even though state would be non-product. 
It is rather unlikely that 
such states exist, yet we have not been able to solve this question.

We now prove a related theorem which follows from the above lemma,
and which will be useful for the following section.  Namely, we show that
using pure states as a resource cannot help when distilling local information.  One can
think of such a process as {\it catalysis} where one uses pure states to produce more pure states
from some shared state.
{\theorem \label{thm:cat}
Local pure ancillas do not help in process of distilling  local information.}\\
\\{\bf Proof.}
Assume that catalysis can help in drawing local information.
Consider a state $\varrho$, which is not the maximally mixed state and the optimal  protocol  of distilling local 
information $P_1$, which  do not use ancillas.   Consider also another protocol $P_2$, in which we  distill information
from some of the copies of state $\varrho$.
Using   $P_1$ and then use the distilled pure states to do catalytic distillation   on the 
rest of copies.
Notice, that we can do this, because we know  from lemma \ref{lemma1} that  we can distill local information  thus 
also pure states from it. If  catalysis is helpful that means that using $P_2$ we are able to obtain more local 
information than in previous protocol. Protocol $P_2$ does not use ancillas and is  better than $P_1$, 
which is optimal. This leads to the required  contradiction.\\

We showed that catalysis is useless for state with nonzero distillable information. It could 
help only in cause of states with pure unlocalisable information, but we know 
from Lemma \ref{lemma1} that such states do not exist.
This  ends the proof.\\
\\
 {\bf Remark.} We know that to do  catalytic distillation  we need pure ancillas. One can notice that states, which we 
want to use in protocol $P_2$ to do catalysis  are not exactly pure. But these  states come from distillation, so 
they are  equal in the limit of many copies to $|0\>^{\otimes rn}$ ($r$ is rate of distillation of local information and 
$n$  is the amount of  copies). This fact assure us that in asymptotic regime of many copies we are able to catalysis.

\section{Can correlations be more quantum than classical?}
\label{sec:super}

The total amount of correlations contained in a bipartite state
is given by the mutual information
\beq
\Im=\SA+\SB-\SAB
\eeq

One can easily see that our quantities for dividing correlations
into ones which behave quantumly ($\D$) and classically ($\Dc$)
satisfy
\beq
\Im=\D+\Dc \s .
\eeq
In other words, the total amount of correlations (given by $\Im$) can
be divided into classical and quantum components.
Now one can ask whether the total correlations $\Im$ can be divided
arbitrarily.  Certainly for pure states, this is not 
the case. For pure states, correlations which behave quantumly cannot
exceed $I/2$.  For pure states
$\psi$, we showed that $\D=\SA$, and thus it is always
the case that $\D(\psi)=\Im/2$.  
For pure states, the quantumness of correlations can never exceed
the classicalness of correlations.  

Now one can ask: Can it be that one has states for which

\begin{equation}
\label{qc>cc1}
\D(\rab)
 > I/2?
\end{equation}
If so, one could think of these states as having {\it super-saturated}
quantum correlations, in that for a given amount of mutual information
$\Im$ they have a greater proportion of correlations which behave
quantumly.  In this sense, one can think of such states as being more
non-local than maximally entangled states.

One way of approach to the above problem is to work with relative 
entropy of entanglement. 
We know that both the relative entropy of entanglement (\(\Er\)), 
with distance taken from separable states, and the von Neumann
entropy (\(S_{AB}\)) are not greater than \(\log_2d\) for \(d \otimes d\) states. Consequently,
one has \(\Er    + S_{AB} \leq  2\log_2 d\).
Can we have the following stronger inequality:
\begin{equation}
\label{Michal}
2\Er    + S_{AB} \mathop{\leq}\limits^{?} 2\log_2 d.
\end{equation}
This is tight for maximally entangled states. Because 
deficit is no smaller than relative entropy of entanglement, 
it follows that if the inequality is violated,
then for some states inequality (\ref{qc>cc1}) is true, 
and we would have the curious phenomenon. On the other side, 
when the inequality is satisfied for all states, 
we would obtain a nice trade-off between entanglement and noise.

In a recent work, Wei \emph{et al.} \cite{Verstraete02-mix-ent} calculated
(for two-qubit states) the maximal 
possible relative entropy of entanglement $\Er$
(as well as other entanglement measures)
 for a given amount of mixedness (quantified 
by the von Neumann entropy). 
Note that the inequality (\ref{Michal})
would generically hold for 
two-qubit states if it is satisfied by these optimal values.
Indeed examining the curves of the above paper, one finds that 
for any two qubit state the inequality is satisfied. 

One can also find that  
for Werner states, and maximally
correlated states, the inequality is satisfied too,
for regularized relative entropy of entanglement.
To see this, the asymptotic relative entropy 
of entanglement (\(E_{r(PPT)}^\infty\)) (with distance taken from states 
with positive partial transpose (PPT))
is known for Werner states (mixture of 
projectors on symmetric and antisymmetric spaces) in \(d\otimes d\) \cite{AEJPVM2001}.
One may check that the relation 
\begin{equation}
\label{Michal2}
2E_{R(PPT)}^\infty + S_{AB} \leq 2\log_2d
\end{equation}
is satisfied for all Werner states in arbitrary dimensions. 
However, note here that the relative entropy of entanglement (from PPT states)
is not additive for Werner states.

For the maximally correlated states, relative entropy of 
entanglement (from PPT states) is known to be additive. Its value is also explicitly
known for all such states in \(d \otimes d\). Via additivity, this would exactly be
equal to its asymptotic relative entropy of entanglement (from PPT states). Precisely, for any state of the form 
\[\varrho_{mc} = \sum_{ij}a_{ij}\left|ii\right\rangle \left\langle jj \right|,\]
we have
\[E_{R(PPT)}^\infty = E_{R(PPT)} = \sum_{i}a_{ii}\log_2a_{ii} - S(\varrho_{mc}).\]
It is easy to check that the relation (\ref{Michal2}) 
is satisfied by any \(\varrho_{mc}\) in \(d \otimes d\).

Thus we haven't found states for which the inequality would be violated 
for regularized relative entropy of entanglement. It remains an open question 
whether the trade-off between noise and entanglement represented by 
inequality is universally true, or whether there exist states,
for which there is more quantum than classical correlations.

\section{Zero-way and one-way subclasses}
\label{sec:zero-one-way}

We now turn to additional measures of quantumness of correlations which
arise when one restricts the communications between Alice and Bob.
In sections \ref{sec:Bell}, \ref{sec:multi}
such restrictions were useful for evaluations of perhaps more basic 
two-way quantities. However they are  
more than just for ease of calculation - we shall
also see that the restricted measures allow one to explore other aspects of
non-locality.  Additionally, there appears to be strong connections between
the deficit and distillation of randomness from shared states.  For example,
it has just been shown in \cite{igor-deficit} that the one-way deficit is 
equal to the mutual information minus the one-way distillable 
randomness \cite{DevetakW03-common}.
%
%
%
%
%
%
%
%

As before, the optimal protocols by which the corresponding local 
informations are obtained amounts to producing 
``classical-like'' states of least entropy by the respective operations. 
As mentioned in section \ref{subsec:main-theorems} the theorems proven there 
apply equally well in these restricted scenarios with suitable modification.

In any protocol of concentrating information  to local form, the parties can stop 
at states of the form 
\ben\label{classical}
\vr_{AB}'=\sum_{ij} p_{ij}\left|i\right\rangle \left\langle i\right| \otimes \left|j\right\rangle \left\langle j\right|.\een
However for two-way scenario, we have argued that one can stop already 
at \pclcor\ states.  When one-way protocols are allowed, it is sufficient for the parties 
to stop at states of the form
\be
\vr_{AB}'=\sum_{i} p_i \left|i\right\rangle \left\langle i\right| \otimes \varrho_i.
\label{eq:one-way-class}
\ee

Finally for zero-way protocols, one has to achieve classical states
(\ref{classical}). 
Consider for example the zero-way protocol  for a state \(\varrho_{AB}\)
 by which \(I_l^{\o}\) is attained.  
Without any classical communication (just by dephasing via an environment),
 Alice and Bob change the state \(\varrho_{AB}\) into  a classical-like state \(\varrho'_{AB}\) (of 
the form given in eq. (\ref{classical})),
so that \(S(\varrho'_A) + S(\varrho'_B)\) is minimized, where \(\varrho'_A\) and \(\varrho'_B\)
are the local density matrices of \(\varrho'_{AB}\).
Note that the parties must concentrate information using classical communication.
But this is only after they have performed all their dephasings.
The situation is therefore like in a Bell-type experiment.
 

Let us now show that \(\Delta^{\o} \)
is an independently useful candidate  for quantum correlations
and can capture interesting aspects of non-locality.
The states that contain no quantum correlations
would be then the ones with \(\Delta^{\o} =0\).
Consider for example the states with eigenbasis (without normalization)
\begin{equation}\label{onewaydistin}
\left|0\right\rangle_A \left|0\right\rangle_B,
\left|0\right\rangle_A \left|1\right\rangle_B,
\left|1\right\rangle_A (\left|0\right\rangle+ \left|1\right\rangle)_B,
\left|1\right\rangle_A (\left|0\right\rangle- \left|1\right\rangle)_B,
\end{equation}
where \(\left|0\right\rangle\) and \(\left|1\right\rangle\)
are the eigenvectors of the Pauli matrix \(\sigma_z\).
Such states are the ones used in the BB84 quantum cryptography protocol \cite{BB84}. 
This set of orthogonal states are distinguishable locally. 
But they are \emph{not} distinguishable by zero-way communication. 
Bob must wait for Alice's measurement result (in the \(\sigma_z\)-basis) 
 to decide whether to perform a measurement
in the \(\sigma_z\)-basis or in the \(\sigma_x\)-basis. 
Therefore a mixture of the states in eq. (\ref{onewaydistin}),
where the mixing probabilities are all different from each other (so that 
the spectrum of the resulting state is non-degenerate), 
would have nonvanishing \(\Delta^{\o}\). This is because an arbitrary dephasing
by Bob on such a mixture, before obtaining
Alice's result would result in no information being extracted from the state (by Bob). 
Consequently there would be an information deficit when trying to extract information locally, because
globally of course all the information is extractable from such a state. 
All the information is also extractable by one-way or two-way communication.
This is contrast to states which have an eigenbasis
\[\left|0\right\rangle_A \left|0\right\rangle_B,
\left|0\right\rangle_A \left|1\right\rangle_B,
\left|1\right\rangle_A \left|0\right\rangle_B,
\left|1\right\rangle_A \left|1\right\rangle_B,\]
for which all the information is extractable from the state locally, by measurement by 
both the parties in the \(\sigma_z\)-basis, without any communication. 

We therefore see that the quantum behavior of correlations could result from the distinctly quantum but ``local'' property
of nonorthogonality. Here we call nonorthogonality a local property, as it does not a priori
require a tensor product structure to manifest itself. It is this nonorthogonality
that manifests itself in a more complex form in the examples of LOCC-indistinguishable 
orthogonal product bases \cite{Bennett-nlwe,BennettUPBI1999,UPBII1999}. 
More generally, it may be the reason for any case of LOCC-indistinguishability 
of orthogonal states
\cite{WH2002, BellPRL, logneg, morenon, ChenLi03-dist}.

\begin{figure}[tbp]
\begingroup%
  \makeatletter%
  \newcommand{\GNUPLOTspecial}{%
    \@sanitize\catcode`\%=14\relax\special}%
  \setlength{\unitlength}{0.1bp}%
\begin{picture}(2339,1728)(0,0)%
\includegraphics{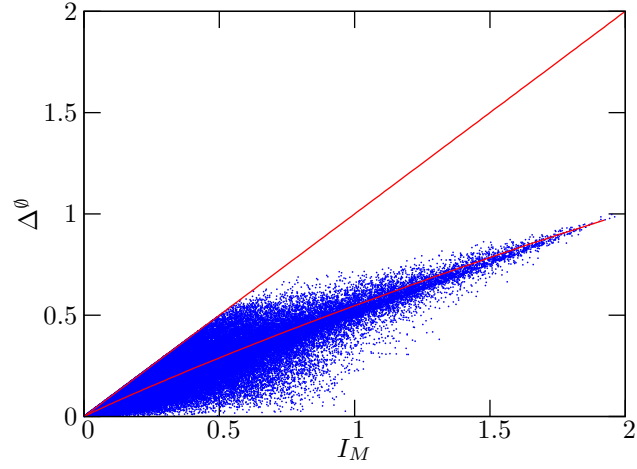}
\fontsize{10}{\baselineskip}\selectfont
\put(1245,25){\makebox(0,0){$I_M$}}%
\put(50,914){%
\makebox(0,0)[b]{\shortstack{$\Delta^\emptyset$}}%
}%
\put(2265,100){\makebox(0,0){ 2}}%
\put(1755,100){\makebox(0,0){ 1.5}}%
\put(1245,100){\makebox(0,0){ 1}}%
\put(735,100){\makebox(0,0){ 0.5}}%
\put(225,100){\makebox(0,0){ 0}}%
\put(200,1678){\makebox(0,0)[r]{ 2}}%
\put(200,1296){\makebox(0,0)[r]{ 1.5}}%
\put(200,914){\makebox(0,0)[r]{ 1}}%
\put(200,532){\makebox(0,0)[r]{ 0.5}}%
\put(200,150){\makebox(0,0)[r]{ 0}}%
\end{picture}%
\endgroup
\caption{Zero-way deficit is plotted versus mutual information for 100 000 
random two qubit states. The upper line stands for $\D^\o=I_M$.
The lower line denotes isotropic states of eq. (\ref{eq:iso}).
Two regimes are evident: in first regime, 
there are states, for which deficit is almost equal to mutual information.} 
\label{fig:lukasz}
\end{figure}

An interesting issue is relation between $\D^\o$ and mutual information. 
In section \ref{sec:super} we have asked a question whether 
there exist states for which $\D$ would be more than half of 
mutual information. The same question can be asked 
in the case of one-way and zero-way deficits. 
\L{}ukasz Pankowski has performed numerical simulations to evaluate 
$\D^\o$ versus mutual information. The results 
are presented on figure \ref{fig:lukasz}. Surprisingly,
there are states, for which the deficit is almost equal to mutual information.
Thus the measurement destroys almost all correlations! The quantum correlations 
do not imply classical correlations (see  \cite{DiVincenzo-locking} in this context). 

With respect to the pure states considered in Section \ref{sec:pure},
it is easy to see that $\D$ is also equal to $\Delta^{\rightarrow}$.
This is also true for single copies of single qubit states, due to Theorem \ref{thm:Nipb-Il}.

\subsection{Expression  for one-way $\Delta$}


In this subsection we consider the expression for the one-way deficit
%
%
In the case when only one-way communication is allowed 
between the parties, the only thing that  Alice and Bob can do, is that Alice dephases 
her part in come basis, and then sends her part to Bob. 
Dephasing transforms the state as 
\[
\varrho_{AB} \to \varrho'_{AB}=\sum_i P_i \otimes I \varrho_{AB} 
P_i \otimes I = \sum_i p_i \left|i\right\rangle \left\langle i \right| \otimes \varrho_B^i
\]
where \(\left\{P_i = \left|i\right\rangle \left\langle i \right| \right\}\) 
forms a set of orthogonal one-dimensional projectors on the 
Hilbert space of Alice's part of \(\varrho_{AB}\) and  $p_i$ are probabilities of the corresponding outcomes 
which Alice would obtain
if she performed measurements with the same $P_i$'s 
rather than dephasing, while $\varrho_B^i$ is the state that Bob 
would obtain conditionally on measurement outcome \(\left|i\right\rangle\). 
Thus
\ben
&&p_i=\tr (\varrho_{AB}P_i\otimes I ) \\ \nonumber
&& \varrho_B^i  = {1\over p_i}
\tr_A (P_i\otimes I \varrho_{AB} P_i\otimes I) 
\een 

The process of sending does not change the form of the state, 
so that the entropy of the final state at Bob
is 
\[
S(\varrho'_{AB})= S(\varrho_A') + \sum_ip_i S(\varrho_B^i)
\]
where $\varrho_A'=\sum_ip_i \left|i\right\rangle \left\langle i \right| $ is the reduced 
density matrix of the \(A\)-part
of $\varrho'_{AB}$. So finally $I_l^\to$  takes the form 
\[
I_l^\to=n_{AB} - \inf_{\{P_i\}} (S(\varrho_A') +  \sum_ip_i S(\varrho_B^i))
\]
and correspondingly
\be
\label{one-waydelta}
\Delta_l^\to=\inf_{\{P_i\}} (S(\varrho_A') +  \sum_ip_i S(\varrho_B^i)) - S(\varrho_{AB}).
\ee

Just as we showed that $\D$ was equal to the relative
entropy distance to \pclcor\ states, one can also write $\D^\rightarrow$ and $\D^\o$ as
the minimum relative entropy distance to the set of states $\cal{S}^\rightarrow$ and $\cal{S}^\o$
which can be created reversibly under the one-way and zero way classes of operations.

\section{Relationship with other measures of quantumness of correlations}
\label{sec:discord}


Let us now compare the deficit with other measures of quantumness of
correlations, in particular the quantum discord
\cite{Zurek-discord,Zurek01-discord}. The latter is defined formally, as the
difference of two classically equivalent expressions for the
mutual information, applied to quantum systems (taken to be a
measuring apparatus and system).
It was defined with respect to a measurement $A_{\cal M}$ (either a projective
one, or a POVM (Positive Operator Valued Measure) performed on the
apparatus {\cal A}.  One then defines the
{\it discord} $\delta(A_{{\cal M}}|B)$
with respect to this measurement
that results with
probabilities $p_{i}$ in joint  states
$\varrho_{AB}'^{i}= (|\psi_{i}\rangle \langle \psi_{i}|)^A \otimes
\varrho_{i}^B$. The discord is defined as 
\beq
\delta(A_{{\cal M}}|B)=H(\{ p_{i} \})
+ \sum_{i}p_{i}S(\varrho_{i}) - S(\varrho_{AB})
\s .
\eeq

The relationship between $\delta(A_{{\cal M}}|B)$ and $\D^\rightarrow$
(defined on single copies)
was recently shown in \cite{Zurek03-demons} where it was shown that
the discord also has the interpretation of extraction of work by a demon,
if one minimizes $\delta(A_{{\cal M}}|B)$ over all possible
measurements $A_{{\cal M}}$.   
Care however must be taken, since with the definition of discord 
there is no cost associated to pure states which are used in a POVM.
Therefore, we note here that the relationship between the discord 
and  $\D^\rightarrow$ only applies if one optimizes the discord over
von Neumann measurements, and disallows POVMs.  

Finally, let us provide two explicit examples of cases where two way
communication is more powerful than 1-way communication.  I.e
one has the strict inequality
$\Delta^{\leftrightarrow}>\Delta^{\rightarrow}=\inf_{A_{{\cal M}}\in
  PV meas}\delta(A_{{\cal M}}|B) $ 
%

To this aim  consider  the basis related to the sausage states of \cite{Bennett-nlwe}
and which has been 
analyzed in \cite{compl}:
 \begin{equation}
  \begin{array}{lll}
    \,&{\mbox \small A}\ &
    {\mbox \small B}\\
    \psi_1=&|0+1\rangle&|2\rangle\\
    \psi_2=&|0-1\rangle&|2\rangle\\
    \psi_3=&|0\rangle&|0+1\rangle\\
    \psi_4=&|0\rangle&|0-1\rangle\\
    \psi_5=&|1+2\rangle&|0\rangle\\
    \psi_6=&|1-2\rangle&|0\rangle\\
    \psi_7=&|1\rangle&|1\rangle\\
    \psi_{8}=&|2\rangle&|2\rangle\\
    \psi_{9}=&|2\rangle&|1\rangle\\
  \end{array}
  \label{9modifiedstates}
\end{equation}
Consider now any bipartite $3 \otimes 3$  state
$\varrho_{two-way}$ that is diagonal  in the above basis,  but has  nondegenerate spectrum.
It is relatively easy to provide a two-way protocol  that distinguishes 
vectors (\ref{9modifiedstates}) without  
destroying  them  (see \cite{compl}).
Hence $\Delta^{\leftrightarrow}$ vanishes. Evidently $\varrho_{two-way}$ is not of the form 
$\sum_{i=1}^{3}|\phi_{i}\rangle \langle \phi_{i}| \otimes \varrho_{i}$ 
with orthogonal $\phi_{i}$, since there are no three eigenvectors  among  
(\ref{9modifiedstates}) that have the same component on Alice's side.
So both $\Delta^{\rightarrow}$ and  discord  are strictly positive for this state.  
Thus Maxwell's demon which communicate in both directions are more powerful than
demon's who can only communicate in one direction.

Another simple example is to take states which have zero optimized discord or one-way deficit
\beq
\rho_\rightarrow=\sum_i p_i \ket{i}_A\bra{i}_A \otimes \rho^i_B \s\s, 
\rho_\leftarrow=\sum_j p_j \rho^j_A \otimes \ket{j}_B\bra{j}_B 
\eeq
but in different directions of communication.  Then take them each to be on orthogonal
Hilbert spaces, and mix.  Such a state will have $\D^\both=0$ since both parties can just
project on the two orthogonal Hilbert spaces to determine whether they hold
$\rho_\rightarrow$ or $\rho_\leftarrow$ and then the appropriate party can send her
state down the channel.  On the other hand,  one-way communication will be sufficient
to completely localize one of the states but not always both.

\section{Relation with measures of classical correlation}
\label{sec:classical}

In this section  we shall analyze the relation of the classical deficit \cite{compl}
to already known measures of classical  correlations.
It happens that both zero-way and one-way deficit have 
their ''counterparts'' in such measures. There is no known 
analog,  however,   for  two-way deficit.

Let us recall that just as the quantum deficit was defined as 
\[
\D=I-\Il
\]
One can think of it as describing how much better Alice and Bob can do under 
closed operations (CO) if they are given a quantum channel instead of 
the classical channel.  Because it feels the difference
between the quantum and classical channel, it tells us about the quantumness of correlations.
Likewise, the classical deficit is given by
\[
\Dc=\Il-\ILO
\]
It tells us how much better two parties can do at localizing information if, instead of having no
access to a channel i.e. closed {\it local} operations, 
they have access to a classical channel.
Because the added resource is a classical channel, it shows how much 
better the parties can do by exploiting a classical channel.

One can verify that $\Dc$ and $\D$ add up to the quantum mutual information 
$I_M(\varrho_{AB})=S(\varrho_A)+S(\varrho_A)-S(\varrho_{AB})$.
Thus
\[
\Delta_{cl} = I_M - \Delta
\]
More explicitly we have (cf eq. (\ref{twowaydel}))
\beq
\Dcl(\varrho_{AB})=S(\varrho_A) + S(\varrho_B) - \inf_{CLOCC}(S(\varrho'_A) + S(\varrho'_B))
\nonumber
\eeq
i.e. $\Dcl$ is the optimal decrease of local entropies 
by means of CLOCC.


\subsection{One-way measures}
\label{subsec:cl-cor-one-way}
Corresponding to the measure of quantumness of correlation under
one-way classical communication (from Alice to Bob) (\(\Delta^\rightarrow\)), 
given by eq. (\ref{one-waydelta}), 
we could have the 
 following formula for classical correlation:
\be
\label{one-wayclass}
\begin{array}{rcl}
&&\Dcl^\to(\varrho_{AB}) \\
&=& \sup_{P_i} [\{S(\varrho_A) - S(\varrho'_A)\} 
+ \{S(\varrho_B) - \sum_i p_i 
S(\varrho_B^i)\}] \\
& \equiv &   \sup_{P_i} [{\delta}S(A) + {\delta}S(B|A)].
\end{array}
\ee
Note that the supremum is taken over all local dephasings on Alice's side.
Although we optimize over projection measurements, one can effectively
include POVM's 
by including all
the required ancillas from the start.
Remarkably, it has been shown \cite{igor-deficit} that POVM's need not be considered
when one goes to to the limit of many copies.

In eq. (\ref{one-wayclass}), we have distinguished two terms. The second term
 \[{\delta}S(B|A) = S(\varrho_B) - \sum_i p_i 
S(\varrho_B^i)\]
shows the decrease of Bob's entropy after Alice's measurement. 
The first one 
\[
{\delta}S(A) = S(\varrho_A) - S(\varrho'_A)
\]
denotes the cost of this process on Alice side,
and is non-positive. It is zero only if Alice measures 
in the eigenbasis of her local density matrix \(\varrho_A\). 

The expression for \(\Dcl^\to\) is very similar to the measure of 
classical correlation introduced by Henderson and Vedral 
\cite{HendersonVedral}: 
\be
\label{HendersenVedral}
\Dhv=\sup_{P_i}( S(\varrho_B) - \sum_i p_i 
S(\varrho_B^i) ).
\ee
Originally the supremum was taken over POVMs, 
but as mentioned we 
take the state acting already on a suitably larger Hilbert space, unless 
stated otherwise explicitly.

The difference between the Henderson-Vedral classical correlation measure 
and one given in eq. (\ref{one-wayclass}) is that the former does not include Alice's 
entropic cost \({\delta}S(A)\)
of performing dephasing. 
Hence in general,
\[\Delta_{cl}^{\rightarrow} \leq \Delta_{HV}.\]
In the asymptotic limit of many copies, one has equality \cite{igor-deficit}.
Actually in \cite{igor-deficit} it was shown that 
regularized one-way classical deficit is equal to another operational 
measures of classical correlations: {\it distillable common randomness} introduced 
in \cite{DevetakW03-common}. The latter is in turn equal 
to regularized Henderson-Vedral measure. It is interesting that 
$\Delta_{cl}^{\rightarrow}$ without regularization, although seems to be an important 
characteristics of classical correlations, does not meet 
a basic requirement for being a measure of classical correlations: it is not monotonous 
under local operations \cite{SynakH04-deltacl}. Thus regularization 
plays here a role of monotonization. There is interesting question 
what happens with two-way classical deficit after regularization.  

\subsection{Additivity of one-way quantum and classical deficits 
for Bell diagonal states}
Here we will prove the fact mentioned in section \ref{subsec:Bell-asymptotic},
that  the one-way deficits are additive, and that borrowing pure qubits 
does not help for Bell diagonal states. 
First of all in \cite{IBMHor2002} it was shown that a measure 
of classical correlations $C_{HV}$ is additive for Bell 
diagonal states. Let us recall the argument, as it will be useful for 
making connection with classical deficit.
For a Bell diagonal state $\rho$, consider a related channel $\Lambda$  
(i.e. such channel that $(I\ot \Lambda) (|\phi\>^+\<\phi^+|)=\vr$).
The maximum output Holevo function over all input ensembles, denoted by $\chi^*(\Lambda)$ 
is, for general channels, no smaller than $C_{HV}$. They are equal,
if the density matrix of ensemble attaining $\chi^*$ is equal to $\vr_A$.
In the case of Bell diagonal states, we have $\vr_A=I/2$, 
and it turns out that the optimal ensemble for corresponding 
channels consists of two orthogonal states, hence gives rise to the 
same matrix.  King \cite{King01-qubit} has shown that $\chi^*$ is additive for 
channels coming from the Bell diagonal states. 
>From this and from the fact that, in general, $\chi^*\geq C_{HV}$
one gets that for Bell diagonal states 
 $C_{HV}$ for many copies is also equal to $\chi^*$ 
for many copies of corresponding channels. This proves that $C_{HV}$ 
must be additive. 

Now, let us make connection with classical deficit.
As discussed in \cite{SynakH04-deltacl}, if $\chi^*$ is attained on such 
ensemble that its density matrix  is equal to $\vr_A$, then 
by looking at ensemble maximizing $\chi$, 
one can tell something about measurements that attain $C_{HV}$.
Namely, when the ensemble is orthogonal, then one attains 
$C_{HV}$ by measurement in eigenbasis of $\vr_A$. 
Now, it is obvious from eq. (\ref{one-wayclass}) and discussion thereafter,
that in the latter case $C_{HV}$ is actually {\it equal} to classical deficit,
as they differ from  one another only by entropy production
during Alice's measurement, which vanishes, if it is done in eigenbasis. 
Since $C_{HV}$ 
is additive, then for many copies it is again attained 
by measurement in orthogonal basis that is eigenbasis of Alice's subsystem. 
Thus classical deficit for many copies is also not less than $C_{HV}$,
and it by eq. (\ref{one-wayclass}) cannot be greater. 

Thus for Bell diagonal states the deficit is equal to $C_{HV}$ 
and it is additive. Moreover, since the measurement was von Neumann one,
the deficit is attained without using POVMs. This means that 
additional pure ancillas do not help. 

So far we have talked about classical deficit. Now,  since quantum and classical 
deficit add up to mutual information  which is additive, it follows that quantum deficit 
is additive too. Also, since borrowing local qubits does not increase 
classical deficit, it cannot decrease quantum deficit. 

\subsection{Zero-way measures}

Let us now consider measures of classical 
correlations under no classical communication, $\Dcl^{\o}$.
Again, this is taken to mean that the parties are not allowed to communicate
before making measurements, but can do so afterward in order to concentrate the classical
records. 
The information deficit under no classical communication, $\Delta^{\o}$, 
is given by 
\[
\Delta^{\o}= S(\varrho_{AB})  - S(\varrho'_{AB})
\] 
where $S(\varrho'_{AB})$ is the von Neumann entropy of the optimal final state $\varrho'_{AB}$ 
(which is classical-like), and was obtained 
by local complete measurements, without  classical communication.
We then obtain 
\[
\begin{array}{rcl}
&& \Dcl^{\o} = S(\varrho_A) + S(\varrho_B)  -  S(\varrho'_{AB}) \\
&=& \{S(\varrho_A)-S(\varrho'_A)\} + \{S(\varrho_B)- S(\varrho'_B)\} \\
&& +  \{S(\varrho'_A) +S(\varrho'_B) - 
S(\varrho'_{AB})\} \\
& \equiv & {\delta}S(A) + {\delta}S(B) + I_M(\rho')  
\end{array} 
\] 
We have three terms here: the last one 
\[I_M(\rho') = S(\varrho'_A) +S(\varrho'_B) - 
S(\varrho'_{AB})\] 
is the 
classical mutual information of the final state, 
while the first two, \({\delta}S(A)= S(\varrho_A)-S(\varrho'_A)\) and 
\({\delta}S(B) = S(\varrho_B)- S(\varrho'_B)\), denote 
respectively the local entropic 
costs of the process at the respective sides. 
We therefore have a trade-off similar to 
that in the one-way case.  And again  there was defined a  
classical correlation measure  \cite{IBMHor2002}
which consists only of the last term of our quantity
\be
C^{\o}= \sup_{P_i} I_M(\varrho')
\label{eq:ibm-cl}
\ee
where $\varrho'$ is obtained out of $\varrho$ by local complete
measurements. Again the original definition of \(C^{\o}\) involved POVMs, but 
as we have suitably increased our Hilbert space from the very beginning, we need not do so.

\section{Complementarity features of information in distributed quantum systems}
\label{sec:compl}

Bohr was the first who recognized a fundamental feature of 
quantum formalism - complementarity between incompatible observables. 
Complementarity was not explicitly related to 
entanglement, now regarded as an important quantum information resource.
Namely, Bohr's complementarity  concerned 
mutually exclusive quantum phenomena associated with a {\it single} system 
and observed under different experimental arrangements.

Let us comment on complementarity in the case of composite systems and Bohr complementarity. 
Roughly speaking, the latter says that one cannot access the properties of the systems 
necessary to describe it by one measurement. The rule is formulated for single quantum 
systems and is a consequence of noncommutativity.

On the other hand we know that one can also divide the properties of the system 
into local and nonlocal ones, and they are complementary with each other too \cite{compl}.
For example, one can perform measurement in Bell basis or in standard product basis.
However one cannot perform those measurements simultaneously. 
In other words one cannot access global and local properties of the system 
(see also \cite{Jakob-compl} in this context).

The latter phenomenon is not merely a consequence of Bohr's complementarity.
Indeed, if the only allowable states of composite systems were the classically correlated 
states:
\be
\rho=\sum_{ij} p_{ij} |e_i\> |f_j\> \<e_i| \<f_j| 
\ee
then maximal information about the total system would be available 
through measurements on subsystems. Global measurements would not access 
any further knowledge about properties of the system. 
On the other hand, Bohr complementarity would still hold, in the sense 
that one cannot access all properties of the system in one measurement. 

Thus we see that the local-nonlocal complementarity \cite{compl} is a consequence 
of two distinct phenomena: {\it noncommutativity
and existence of entanglement (or quantum correlations)}.
So not only is there noncommutativity, 
but there is too much of it, so that it affects also relations between local and nonlocal 
informational contents.


In distributed systems one usually imposes constraints by 
allowing operations that can be done solely by 
classical communication and local operations.
It turns out that in such situation there also arises an interesting complementarity.
Namely, in \cite{compl} we considered two tasks: localizing information 
(which we have presented in this paper) and sending quantum information (e.g. teleportation),
performed simultaneously. 
It was shown that for a fixed protocol $\pcal$, the 
rates of those two tasks obey the following relation
\be
I_l(\pcal,\rho)+ Q(\pcal,\rho)\leq I_l(\rho)
\ee
where $I_l(\pcal,\rho)$ is the amount of information localised by 
the protocol $\pcal$ and $Q(\pcal,\rho)$ is the amount of qubits transmitted 
by the protocol.

For example, for the singlet state, the total informational contents 
is equal to total correlation contents  and amounts to two bits. 
The right hand side of the inequality is equal to $1$. 
This number $2$ in the light of the above complementarity 
we can interpret as follows: $2$ is equal not to $1$ plus $1$ but 
it is equal to $1$ {\it or} $1$.
One can either draw one bit of local information (classical correlations) or 
teleport one qubit (quantum correlations), however we cannot access both 
bits. 

One can see that this phenomenon is connected with 
above-mentioned Bohr complementarity 
for distributed systems: for the task 
of teleportation, Alice makes a Bell measurement on her part 
of the singlet and the unknown state to be sent, while to localize information,
she measures only the half of singlets. Interestingly, 
as far as those two exclusive measurements are concerned,
the "local versus nonlocal" complementarity occurs within Alice laboratory,
while it results in complementarity between tasks 
that refer to local-nonlocal properties of systems belonging to Alice and 
Bob.

The above inequality suggests an interesting problem: to find the trade-off 
curves for performances of teleportation and localizing information 
of a given state. In particular, an interesting question is whether 
there exist states for which if we teleport the amount of 
qubits equal to distillable entanglement, one not only would 
not localize any information, but would need to spend some 
additional pure states (see \cite{phase} in this context).

\section{Discussion and open questions}
\label{sec:conclusions}
In conclusion we have developed the quantum information processing paradigm 
which involves local information as a natural resource in the context 
class of CLOCC operations. We have presented proof that the central quantity of the paradigm, 
quantum information deficit is  bounded  from above by the relative entropy distance from 
the set of pseudo-classically correlated states. 
We showed how the paradigm allows one to define thermodynamical cost of erasure of entanglement:
entropy production necessary to make state separable by CLOCC operations.
We proved that the cost is no smaller than relative entropy of entanglement. 
Since the cost is no greater than the deficit, we have obtained that the 
deficit is no smaller than relative entropy of entanglement. 
This in turn implies  that {\it every entangled state}
exhibits informational nonlocality. 

We have also found that the paradigm offers a new method of analysis 
of correlations of multipartite states. The most nonlocal state  from 
this point of view (we call it informationally nonlocal)
would be the one for which one has to produce the largest entropy  
while converting it into classical states. It turned out that according 
to such a criterion, the Aharonov state is much more nonlocal 
than GHZ one. The nonlocality that can be probed by our methods 
is one that is not caught by Bell's inequalities, since 
we have found that also separable states can exhibit nonzero deficit.
Rather, it has much in common with nonlocality without entanglement, 
that was found for ensembles of states \cite{Bennett-nlwe}.
Thus our nonlocality is not identical with entanglement. 
As a matter  of fact it is a wider notion. 

The information deficit has then some peculiar properties. Since 
it is not an entanglement measure, it can increase under local operations. 
It is not unreasonable: Local operations may destroy a
local property, and make it impossible to carry out some action
by separated parties, while when the parties meet, 
the action may still be achievable. This curious behavior 
of quantum states may be attributed to the fact 
that even for separable states, when they are mixtures 
of nonorthogonal states, we cannot ascribe to the subsystems
local properties (this may have some connection with the Kochen-Specker theorem). 

The paradigm developed in this paper opens many important questions. 
Here are some of them. 
\bei
\item {\it Are "noncommuting-choice  protocols" better in 
localizing information?} This is the major problem in 
the paradigm of localizing information by CLOCC operations.

\item {\it Is the quantum deficit equal to relative entropy 
distance to \pclcor\ states?} This question would be answered positively,
if the noncommuting-choice  protocols do not help. 

\item {\it Is regularized deficit still nonzero for all 
entangled states?} For regularized deficit we have lower bound 
given by regularized relative entropy distance. However 
we do not know if for any entangled state 
the latter is nonzero.

\item {\it Is deficit for multiparty pure states 
equal to relative entropy of entanglement?}
For bipartite states it was proven that the deficit is equal 
to entanglement. For multiparty case it is also true for Schmidt 
decomposable states. It is an open problem whether it is true in general. 
The same question  can be asked  for regularized deficit. Is it 
equal to regularized $E_r$ for multipartite pure states?

\item {\it Is two-way classical deficit a legitimate measure of 
classical correlations?} The classical deficit definitely is important 
quantity describing some aspects of classical correlations. However 
there is a question, whether it can be used to quantify them. To this end, 
it should not increase under local operations \cite{HendersonVedral}.
For one way case, the classical deficit is not monotonous 
under local operations as shown in \cite{SynakH04-deltacl}.
Yet it turns out that after regularization, the monotonicity is regained 
\cite{igor-deficit}, because regularized one-way classical deficit is equal 
to one-way distillable common randomness of \cite{DevetakW03-common}. 
Can two-way classical deficit be also monotonous after regularization? This 
is connected with the next question:

\item {\it Is classical two-way deficit equal to two-way 
distillable common randomness \cite{DevetakW03-common}?}

\item {\it Is relative entropy of entanglement 
the thermodynamical cost of erasure of entanglement?}
We have shown that  the cost is bounded from below by relative entropy of entanglement.
If there is equality, relative entropy of entanglement 
would acquire operational status:
it would  be interpreted as thermodynamical cost of {\it erasure of entanglement}.

\item {\it What is the relation between deficit and  mutual information?}
We have shown that if a trade-off inequality for $E_r$ (\ref{Michal}) 
would be violated, then quantum deficit would be more than the classical 
deficit for some states.   
We have also touched on this question by analysis of zero-way deficit versus mutual information. 
Preliminary results suggest that there  is very interesting phenomenon
while going from quantum to classical states via local measurements: for some states 
before measurement there are large correlations quantified by mutual information, while 
after measurement, the remaining amount of information 
is equal almost exclusively to initial local information. 
This means that for some states, even optimal measurement may destroy
most of information contained in correlations.  The question can be recast 
in the following way: how small can be the classical deficit versus 
mutual information? 

In \cite{DiVincenzo-locking}  measure of classical 
correlations (\ref{eq:ibm-cl}) closely related to 
zero-way deficit  was compared with mutual information. 
The authors showed that when this measure is smaller than $\epsilon$ 
then mutual information is smaller than $\epsilon\, \mbox{poly}(d)$ 
where $d$ is dimension of the Hilbert space. 
They were however unable to improve the factor to be of order of $\log d$. 
This means that most probably there is place for dramatic divergence 
between the two measures of correlations. 
Since deficit can be only smaller from the measure of (\ref{eq:ibm-cl}), 
the effect can be even stronger. All that suggests 
that there may be a large gap between the classical and quantum. 

\item A fundamental open problem, or rather program is to analyze complementarity 
between drawing local information and distilling singlets initiated in 
\cite{compl}. In the latter paper, the two tasks: drawing local information 
and teleporting qubits were treated as complementary ones. One obtains  
trade-offs, if one wants to perform those tasks simultaneously. An open question 
is whether distilling singlets can lead to negative amount of local information gained,
i.e. whether in process of distillation we have to use up local pure qubits 
rather than we gain them \cite{phase}.  
Moreover one can define  the following quantity: maximal amount of 
pure qubits one can draw  by CLOCC from a given state \cite{SynakHH04}. Note that 
here we do not speak about local qubits. Thus for example, singlet is already pure 
and needs no action. Due to reversibility 
in entanglement transformations for pure bipartite states \cite{BBPS1996}, the question 
in fact reduced to the problem of  drawing simultaneously singlets and local pure qubits. 

\item An interesting question arises in the context of 
\cite{GroismanPW04}. There the authors probe correlations 
by applying random local unitaries  to transform  the state to product 
or separable form, using the smallest number of unitaries. 
This method allows to define not only quantum correlations 
but also total correlations in terms of entropy production 
while reaching some set of states. It differs from our approach 
in that the authors do not use classical communication in an essential way (it cannot help).
Therefore a natural application of their method is to probe total correlations.
This allows them to give a fresh, operational meaning to the quantum mutual information --
it is the entropy production needed to bring a quantum state into product form.
Our method could be applied in a similar way - one tries to bring a state into product
form using CLOCC but without the classical communication (i.e. CLO)
Then one finds that the entropy production (i.e. deficit to
product states $\Delta^{CLO}_{prod}$) is equal to $I(\rho_{AB})$.  This can be seen
simply from the fact that the optimal protocol is for one party to locally compress her state
and then to dephase in the eigenbasis of the compressed state.  She then dephases in a basis complementary
to the eigenbasis.  The latter measurement completely destroys all correlations between
$A$ and $B$.  Since the initial entropy was $S(\rho_AB)$ and the final entropy is $S(A) + S(B)$,
the deficit and hence entropy production, is $I(\rho_{AB})$.  Just as the relative entropy distance
to some set of states (\pclcor, and separable states) played a crucial role in the case of $\Delta$ and
$\Delta_{sep}$, here, the relative entropy distance to product states plays the crucial role,
and is equal to the quantum mutual information.  

It is rather amusing that this gives the same answer as the method used in \cite{GroismanPW04},
since in our cases, Alice performs her measurement without any knowledge of the density matrix
of Bob, while in \cite{GroismanPW04}, she must use this information.  Furthermore, the number of unitaries
which would be needed to perform the dephasing in our case, is $S(A)^2$, 
far greater than the optimal number found in \cite{GroismanPW04}.  Understanding in greater
detail why these two methods give the same answer might be an interested avenue of further
research.  It is also interesting to compare how one divides the total correlations into quantum
and classical ones.  For example, in the case of the singlet, \cite{GroismanPW04} interpret
the two bits of mutual information as requiring one bit of noise to destroy the entanglement,
and one bit of noise required to destroy the secret correlations.  In \cite{compl} we interpreted
the two bits in terms of one use of a quantum channel, {\it or} one bit of local information.

In the case of destroying correlations due to entanglement, our method uses
classical communication in an essential way, therefore on the surface, it appears
to naturally encode the notion of entanglement whose definition relies on the class of LOCC.
For pure states the authors of \cite{GroismanPW04} also obtain entanglement, 
as in this case communication is not needed to reach the set of separable  
states. It is interesting then to compare what those both approaches 
would produce as far as entropic cost of erasing entanglement 
is concerned. One could expect that our method will show less cost in the case of
erasing entanglement.


\eei

Finally we strongly believe that the present, 
novel paradigm analyzed and developed here
will be helpful as a new rigorous tool in searching for a border
or rather a way of coexistence between 
quantumness and classicality in physical states.
It may also enrich our understanding of quantum 
information processing and its relation 
to other branches of physics like  
thermodynamics and statistics.

\begin{acknowledgments}
We thank \L{}ukasz Pankowski for performing simulations of figure \ref{fig:lukasz},
and Sandu Popescu and Andreas Winter for discussions on their draft of \cite{GroismanPW04}.
We also thank Igor Devetak for discussions on \cite{igor-deficit}.
This paper has been supported by the Polish Ministry of Scientific
Research and Information Technology under the (solicited) grant No.
PBZ-MIN-008/P03/2003 and by EC grants RESQ, Contract No. IST-2001-37559 
and  QUPRODIS, Contract No. IST-2001-38877.
JO also acknowledges the support of the Lady Davis
Trust, and ISF grant 129/00-1 as well as support from the Cambridge-MIT 
Institute.  AS and US also acknowledge support from  the Alexander 
von Humboldt Foundation.
\end{acknowledgments}

\bibliography{../refmich,../refjono,huge}




\end{document}